%% file: main.tex
\documentclass[%
reprint,
amsmath,amssymb,
aps,
prb,
floatfix,
longbibliography
]{revtex4-2}

\usepackage{graphicx}

\usepackage{dcolumn}
\usepackage{bm}

\makeatletter
\newcommand{\hathat}[1]{%
\begingroup%
  \let\macc@kerna\z@%
  \let\macc@kernb\z@%
  \let\macc@nucleus\@empty%
  \hat{\mathchoice%
    {\raisebox{.4ex}{\vphantom{\ensuremath{\displaystyle #1}}}}%
    {\raisebox{.4ex}{\vphantom{\ensuremath{\textstyle #1}}}}%
    {\raisebox{.48ex}{\vphantom{\ensuremath{\scriptstyle #1}}}}%
    {\raisebox{.48ex}{\vphantom{\ensuremath{\scriptscriptstyle #1}}}}%
    \smash{\hat{#1}}}%
\endgroup%
}
\makeatother

\usepackage{acronym}
\newacro{STM}[STM]{scanning tunneling microscopy}
\newacro{EPR}[EPR]{electron paramagnetic resonance}
\newacro{DLT}[DLT]{damping-like torque}
\newacro{FLT}[FLT]{field-like torque}
\newacro{GKSL}[GKSL]{Gorini–Kossakowski–Sudarshan–Lindblad}
\newacro{QD}[QD]{quantum dot}
\usepackage{feynmp-auto}
\usepackage{bbold}
\usepackage{cancel}

\usepackage{lipsum}
\usepackage{comment}
\usepackage{placeins}

\begin{document}

\title{Coherent and Dissipative Spin Torques in Quantum Dots: A Unified Framework for Quantum Spin Dynamics}

\author{Dominic Ruckert}
\author{Stepan Kovarik}
\author{Richard Schlitz}
\author{Mirco Grellmann}
\author{Aishwarya Vishwakarma}
\author{Pietro Gambardella}
\author{Sebastian Stepanow}

\affiliation{Department of Materials, ETH Zurich, 8093 Zurich, Switzerland.}%

\date{\today}

\begin{abstract}
The manipulation of single spins through spin-polarized tunneling opens new routes for quantum control at the atomic scale. We present a theoretical framework describing spin-transfer, spin torques and spin resonance in molecular quantum dots weakly coupled to magnetic electrodes. By deriving a Lindblad master equation from microscopic tunneling processes, we capture both coherent exchange interactions and dissipative spin torque effects within a unified approach. We analyze how charge transport through localized orbitals influences spin dynamics and show that modulating the tunneling rates in time can induce electron spin resonance. This framework is further extended to coupled spin systems, revealing how spin coherence and entanglement respond to local spin torques and highlighting sources of transport-driven decoherence. Our results provide a general model to interpret spin-resolved tunneling experiments and extend classical spin torque concepts into the quantum regime.
\end{abstract}

\maketitle

\section{Introduction}
 Electron transport in quantum systems can be used to sense and manipulate single and coupled spin momenta. In general, the current flow is governed by spin selection rules, enabling the characterization of spin relaxation and coherence times, the identification of spin-orbit and hyperfine interactions, and the operation of spin qubits. These phenomena have been widely studied in theoretical and experimental investigations of \acp{QD}~\cite{camarasa-gomez_spinorbit_2024, khomitsky_spin-flip_2025, paaske_exchange_2010, willke_probing_2018, kovarik_spin_2024, crisan_harnessing_2016, koppens_driven_2006, mcmillan_image_2020, reina-galvez_study_2021}, 
which provide an ideal platform for investigating the dynamics of spins coupled to a dissipative environment~\cite{hanson_spins_2007,crisan_harnessing_2016, koppens_driven_2006, duprez_spin-valley_2024, delteil_observation_2014,raghunandan_initialization_2020}.
Their simple electronic structure allows for a quantum transport description, where many-body interactions can be treated perturbatively within second quantization~\cite{braun_theory_2004, stocker_coherent_2024}. 
As a result, QDs display a wide range of spin phenomena—including coherent precession, relaxation, and decoherence—that can be quantitatively modeled at the single-particle level.

The investigation of driven spin dynamics in QD-electrode systems has revealed two distinct microscopic mechanisms by which spin resonance can be induced. In such systems, the QD's spins are subject to both local magnetic fields from nearby electrodes and exchange coupling arising from tunneling processes. When the exchange interaction is modulated at radio frequencies, it effectively mimics an oscillating transverse magnetic field, thereby driving spin resonance in a manner analogous to conventional electron paramagnetic resonance (EPR). This field-like mechanism is coherent in nature and has been demonstrated in several platforms, including semiconductor QDs~\cite{nowack_coherent_2007,laucht_electrically_2015} and adatoms probed by scanning tunneling microscopy (STM)~\cite{baumann_electron_2015,yang_coherent_2019, seifert_longitudinal_2020,wang_atomic-scale_2023,kovarik_spin_2024}.

A complementary driving mechanism arises when spin-polarized tunneling itself is modulated at the Larmor frequency. In this case, the spin-selective charge transfer exerts a dissipative spin-transfer torque on the localized spin, leading to resonance excitation~\cite{kovarik_spin_2024, shakirov_spin_2019}. While both mechanisms give rise to transport-detected EPR, their origins differ fundamentally: exchange modulation produces a coherent, Hamiltonian-type driving, whereas the dissipative spin transfer torque reflects nonconserving particle processes rooted in spin-dependent tunneling. Crucially, these effects are not restricted to QD and STM junctions, but are generic features of spin-dependent transport, highlighting the close connection between single-spin dynamics and mesoscopic spintronic phenomena.

\begin{table*}[ht]
\caption{\label{tab:spin_torque}Classification of spin torque components corresponding to coherent and dissipative processes and related terminology adopted in spintronics and quantum electronics. 
}
\begin{ruledtabular}
\begin{tabular}{lll}
Spin torque components & Coherent & Dissipative \\
\hline
 & Field-like torque  & Damping-like torque \\
 & Field-like torque & Antidamping-like torque \\
 Spintronics & Field-like torque & Spin transfer torque \\
 & Rashba torque & Slonczewski torque \\
 & Out-of-plane torque & In-plane torque \\
 & Spin-orbit field & Spin Hall torque \\
\hline
 & Spin torque & Spin accumulation \\
 Quantum electronics & Exchange field & Spin accumulation \\
 & $B_1$ & Spin transfer torque \\
\end{tabular}
\end{ruledtabular}
\end{table*}

Effects known from spintronics, such as spin torques~\cite{slonczewski_conductance_1989, stiles_anatomy_2002, ralph_spin_2008, manchon_current_2019} and the Hanle effect~\cite{johnson_interfacial_1985,lou_electrical_2007, raes_spin_2017}, thus also emerge in QDs, albeit in forms that do not yet have a complete theoretical description at the single-spin level~\cite{zaffalon_zero-dimensional_2003, kovarik_spin_2024, kuznetsova_hanle_2013, braun_hanle_2005,mondal_quantum_2019,petrovic_spintronics_2021}.
Spin torques are central to current-induced magnetization switching and oscillations in nanoscale and microscale magnetic structures. They are commonly distinguished into two classes: \textit{field-like torques} (FLT), driving coherent precessional dynamics, and \textit{damping-like torques} (DLT), which act dissipatively to align spins with a spin-polarized current~\cite{slonczewski_conductance_1989, stiles_anatomy_2002, ralph_spin_2008, manchon_current_2019}. While these mechanisms are well-established in ferromagnetic multilayers and magnetic tunnel junctions, their manifestation at the level of single-electron orbitals, where transport is quantum-coherent and fluctuations are pronounced, remains an open question. Molecular \acp{QD} offer a powerful platform to revisit these concepts at the orbital scale, providing tunable control over spin-charge interactions and enabling the study of spin torques in regimes far beyond conventional mesoscopic devices.

Understanding the microscopic origin and the effect of spin torques at the single-spin level requires theoretical models that go beyond closed-system Hamiltonians~\cite{lado_exchange_2017, gunderson_floquet_2021, schnell_high-frequency_2021, verso_dissipation_2010}. In particular, an open quantum system description is essential to capture dissipative effects arising from spin-selective tunneling between QDs and spin-polarized electrodes~\cite{anderson_localized_1961, appelbaum_exchange_1967, ternes_spin_2015, braun_theory_2004, ast_theory_2024, reina-galvez_study_2021, reina_galvez_cotunneling_2019}. 
Recent advances in modeling QDs and molecular junctions have combined microscopic quantum impurity models with open-system techniques, enabling a unified description of both coherent FLT and dissipative DLT. Such frameworks bridge experimental observations of spin dynamics in single atoms and molecules with the fundamental mechanisms underlying spin-transfer effects~\cite{delgado_spin-transfer_2010,shakirov_spin_2019}.
However, in these models, the coherent and dissipative dynamics are hidden in the complex mathematical formulation of the master equations, and it is not directly evident how time-dependent spin torques lead to spin-locking and magnetic resonance.

\begin{figure}
\includegraphics[scale=0.1875]{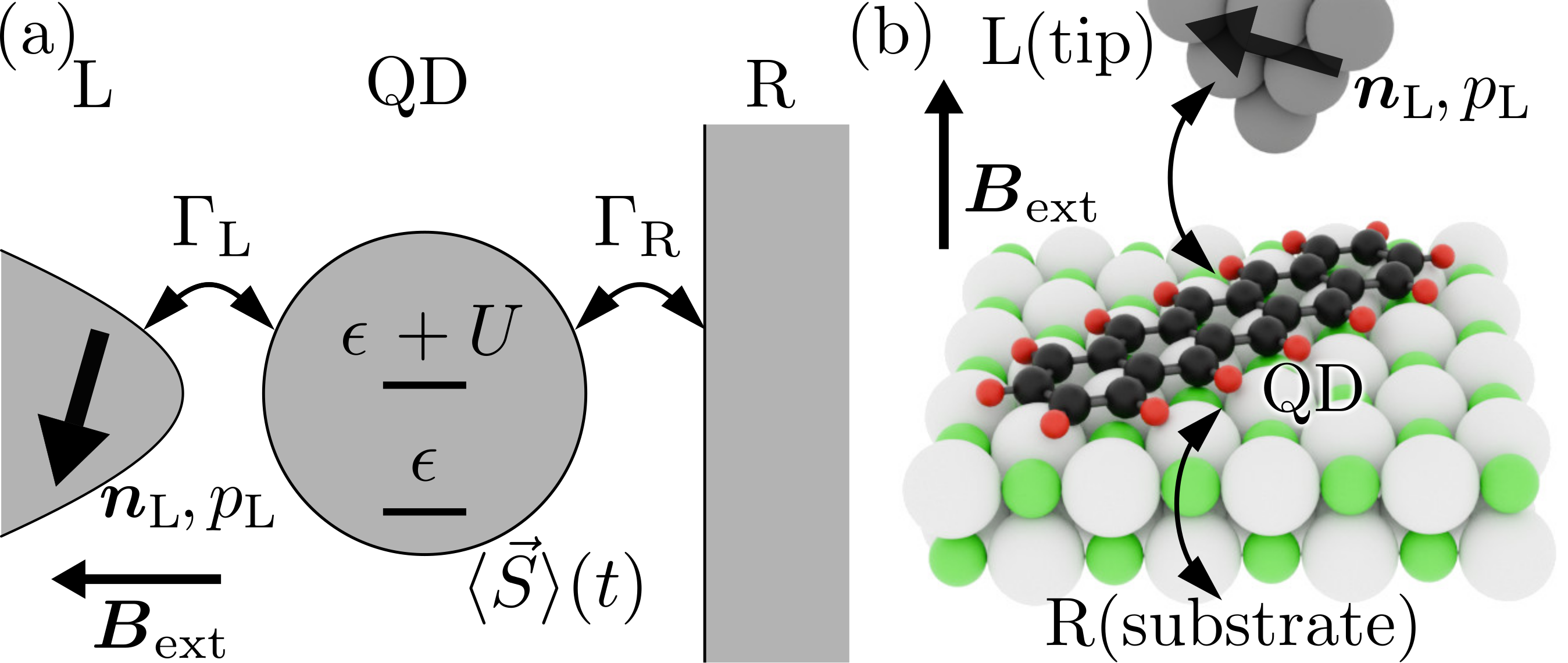}
\caption{\label{fig:epsart} (a) Schematic of a QD weakly coupled to two electrodes where the left electrode is spin-polarized with polarization $p_\mathrm{L}$ along $\boldsymbol{n}_\mathrm{L}$ and the right electrode is non-magnetic. (b) Experimental realization of a molecular QD system using a pentacene molecule in an STM junction that is adsorbed on a thin film of MgO decoupling the molecule form the metallic substrate~\cite{kovarik_spin_2024}. The STM tip serves as a spin-polarized electrode.}
\label{fig:micro}
\end{figure}

In this paper, we employ a density matrix framework that readily allows for distinguishing between coherent and dissipative contributions to spin dynamics in a molecular QD coupled to spin-polarized electrodes. The system is modeled as a weakly coupled open QD bridging a magnetic and a nonmagnetic electrode, as typically realized in an STM junction (Fig.~\ref{fig:epsart})~\cite{braun_theory_2004, reina-galvez_study_2021, busz_hanle_2023,mcmillan_image_2020}. Section~\ref{sec:context} introduces the physics of spin torques in the context of spintronics and QDs. Section~\ref{sec:theory} presents our theoretical framework: we introduce the Hamiltonian in Sec.~\ref{sec:Hamiltonian}, analyze scattering processes and derive real-time equations of motion in Sec.~\ref{sec:4x4} using diagrammatic techniques~\cite{braun_theory_2004}, and reformulate these equations in Lindblad form in Sec.~\ref{sec:Lindblad-me}. This allows us to attribute the emergence of exchange fields and damping-like transfer of spin-angular momentum to coherent and dissipative processes, respectively, and to generalize the model to time-dependent driving relevant for spin-torque-driven \ac{EPR}. 
In Sec.~\ref{sec:currents}, we derive expressions for the charge and spin currents, and in Sec.~\ref{sec:pheno}, we connect the microscopic spin torques in the QD to their mesoscopic analogs.
In Sec.~\ref{sec:DC-Hanle} we apply our model to simulate DC transport in single and coupled QD configurations, including a "sensor--spectator" setup involving an Ising spin or an entangled spin-1/2 pair. We visualize the resulting spin dynamics and transport using a generalized Bloch trajectory representation and explore the loss of entanglement through simultaneous transport in the coupled system. 
Finally, in Sec.~\ref{sec:AC-STT} we address AC-driven dynamics, focusing on the emergence of spin-torque-induced \ac{EPR}, its relation to DC transport features, and the interpretation of \ac{EPR} spectra in the magnetic field domain.

\section{Coherent and dissipative spin torques}\label{sec:context}

A spin torque embodies the transfer of spin angular momentum from "transport" electrons to "local" electrons. In all generality, a spin torque can have three components, two orthogonal to the local spin moment $\boldsymbol{S}$ (or magnetization), and one parallel to it. The two orthogonal components are further distinguished in an energy-conserving FLT component $\sim\boldsymbol{s}\times\boldsymbol{S}$ and a dissipative DLT component $\sim\boldsymbol{S}\times(\boldsymbol{S}\times\boldsymbol{s})$, where $\boldsymbol{s}$ is the spin moment of the traveling electrons~\cite{slonczewski_conductance_1989, stiles_anatomy_2002, ralph_spin_2008, manchon_current_2019}. Both of these components induce a rotation of $\boldsymbol{S}$. In contrast, the longitudinal torque $\sim\boldsymbol{s}\parallel\pm \boldsymbol{S}$ induces an increase (reduction) of $\boldsymbol{S}$ for parallel (antiparallel) alignment of $\boldsymbol{s}$ and $\boldsymbol{S}$~\cite{noel_nonlinear_2025}. We note that the term "torque" in spintronics is used in a generalized sense
to indicate transfer of spin angular momentum. The longitudinal component thus describes spin pumping and relaxation, whereas only the transverse components correspond to proper mechanical torques. The longitudinal torque has a dissipative character and is therefore considered as a DLT in the remainder of this paper.

Although the distinction between torque components is rather straightforward, the discussion of spin torques is often complicated by overlapping terminology and different wording used in the QD, STM, and spintronic literature. In the context of quantum electronics, \textit{spin torque} refers only to the energy-conserving coherent component of the torque, whereas the dissipative component is associated to \textit{spin accumulation}, representing the nonequilibrium spin moment of a QD. In spintronics, instead, spin accumulation refers to differences in the spin-dependent electrochemical potential of electrons on either side of a material's interface, reflecting an imbalance in spin population. Such a spin accumulation can give rise to both coherent FLT and dissipative DLT spin torque components. A \textit{spin transfer torque} (STT) usually refers to both of these torque components when they are induced by a charge current flowing from a reference magnetic layer, acting as spin-polarizer, into a second magnetic layer~\cite{ralph_spin_2008}. A \textit{spin-orbit torque} (SOT) refers to the torque components produced by a charge current flowing into a nonmagnetic layer onto the magnetization of an adjacent magnetic layer~\cite{manchon_current_2019}. STT relies on the transfer of spin momentum between different magnetic systems, whereas SOT exploits charge-spin conversion mediated by the spin-orbit interaction in nonmagnetic systems, as in the spin Hall and Rashba effects. However, both STT and SOT arise from the exchange interaction between a spin current carried by $s$-like conduction electrons and the local magnetization arising from partial occupation of $d$-like orbitals in a magnetic material~\cite{stiles_anatomy_2002,zhang_mechanisms_2002, gambardella_current-induced_2011,amin_spin_2016}.
Therefore, STT and SOT differ only in the origin of the spin current, but not in the microscopic processes attendant to the torque. 

In STM, spin torques emerge when magnetic tips are used to probe magnetic layers~\cite{wiesendanger_spin_2009,balashov_magnon_2006} and single or few-spin systems akin to molecular QDs~\cite{krause_joule_2011,palotas_enhancement_2016,loth_spin-polarized_2010,khajetoorians_current-driven_2013}. Inelastic spin-flip electron tunneling processes, which have been widely studied in the last two-decades as a spectroscopy tool, are essentially the manifestation of a longitudinal spin torque~\cite{heinrich_single-atom_2004, hirjibehedin_spin_2006, loth_controlling_2010, tsukahara_adsorption-induced_2009, mugarza_spin_2011, lorente_efficient_2009,delgado_spin_2010,chen_probing_2008,fernandez-rossier_theory_2009}. More recently, quantum coherent spin control has been achieved by implementation of electron paramagnetic resonance (EPR) in STM~\cite{baumann_electron_2015, yang_coherent_2019, willke_probing_2018, willke_coherent_2021,seifert_single-atom_2020,seifert_longitudinal_2020,veldman_free_2021,steinbrecher_quantifying_2021,kovarik_electron_2022, kovarik_spin_2024}. By tuning the coupling strength between tip, atom, and substrate—either via tip positioning or through the use of decoupling layers, STM experiments can access regimes ranging from strongly hybridized to nearly isolated spins. This tunability allows for systematic studies of spin relaxation, decoherence, and spin-environment interactions. EPR-STM further enables quantum sensing protocols to probe and control single spins on surfaces~\cite{choi_atomic-scale_2017, esat_quantum_2024,del_castillo_theory_2025}. However, a direct connection between EPR, coherent, and dissipative spin torques has been established only recently, showing that virtual and real electron tunneling processes in molecular orbitals govern the relative strength of FLT and DLT components, offering microscopic insight into different spin torque mechanisms~\cite{kovarik_spin_2024,reina-galvez_contrasting_2025}.

In the following sections, we introduce a comprehensive model of coherent and dissipative spin torques in molecular QDs induced by either direct (DC) or alternating currents (AC), which is directly applicable to predict and simulate the current-voltage characteristics of single and coupled spins in a QD or an STM junction. To connect the terminology used in the fields of spintronics and quantum transport, we summarize in Table~\ref{tab:spin_torque} some of the most used and unfortunately confusing torque definitions adopted in the literature. 
Throughout this work, we will refer to the different spin torque components as coherent and dissipative or, equivalently, as FLT and DLT, respectively. 

\section{Microscopic model}\label{sec:theory}
Figure~\ref{fig:micro}(a) provides an overview of the model system considered here, namely a two-level QD coupled to a magnetic electrode. Such a system can be realized in an STM junction including a radical molecule such as anionic pentacene and a magnetic tip, as shown in Fig.~\ref{fig:micro}(b).
We consider a \ac{QD} with many-body basis states $|\hspace{0.15em}\emptyset\hspace{0.15em}\rangle$, $|\!\uparrow\,\rangle$, $|\!\downarrow\,\rangle$, and $|\hspace{0.15em}2\hspace{0.15em}\rangle$ defined as
\begin{align}
    |\hspace{0.15em}\emptyset\hspace{0.15em}\rangle&=|N=0;S=0\rangle\,, \label{eq:def_vac}\\
    |\!\uparrow\,\rangle&=|N=1;S=\tfrac{1}{2},m_s=\tfrac{1}{2}\rangle\,,\label{eq:def_up}\\
    |\!\downarrow\,\rangle&=|N=1;S=\tfrac{1}{2},m_s=-\tfrac{1}{2}\rangle\,,\label{eq:def_dw}\\
    |\hspace{0.15em}2\hspace{0.15em}\rangle&=|N=2;S=0\rangle=\tfrac{\sqrt2}{2}\left(|\!\uparrow\downarrow\rangle-|\!\downarrow\uparrow\rangle\right)\,.\label{eq:def_two}
\end{align}
Here, $N$ denotes the electron occupation number, $S$ the total spin, and $m_s$ the spin projection along the $z$-axis. The external field $B_\text{ext}$ can assume any orientation in the junction geometry. Without loss of generality, we take $B_\text{ext}$ along $z$ (Fig.~\ref{fig:micro}). The states $|\hspace{0.15em}\emptyset\hspace{0.15em}\rangle$ and $|\hspace{0.15em}2\hspace{0.15em}\rangle$, corresponding to the uncharged and doubly charged molecule, respectively, are spin singlets ($S=0$), whereas  $|\!\uparrow\,\rangle$ and  $|\!\downarrow\,\rangle$ form a spin doublet ($S=\tfrac{1}{2}$). 

The \ac{QD} is tunnel-coupled to a left and a right electrode, with coupling strengths $\Gamma_\mathrm{L}$ and $\Gamma_\mathrm{R}$, respectively.
As illustrated in Fig.~\ref{fig:micro}(a), the setup is applicable to both semiconductor \acp{QD}~\cite{braun_theory_2004} and molecular tunnel junctions, where tunnel barriers separate the dot from both electrodes~\cite{reina-galvez_study_2021}.  In the latter case, a double barrier configuration can be realized in an STM setup, where a molecule is adsorbed on an insulating decoupling layer atop a metallic substrate, as shown in Fig.~\ref{fig:micro}(b). This results in a weak coupling of the molecular \ac{QD} to both the \ac{STM} tip and the substrate.
In the following, we consider a specific configuration where the left electrode (STM tip) is spin-polarized, while the right electrode (metallic substrate) is nonmagnetic. 

\subsection{Hamiltonian}\label{sec:Hamiltonian}
A microscopic description of the \ac{QD} coupled to its environment is given by an extended Anderson impurity model~\cite{anderson_localized_1961}, defined by
\begin{align}
\hat{H}&=\hat{H}_0 + \hat{H}_{\text{Z}} +\sum_{\alpha=\mathrm{L},\mathrm{R}} \hat{H}_{\alpha} +  \hat{V}_\alpha\,,\label{eq:H} 
\end{align}
where
\begin{align*}
\hat{H}_0&= \epsilon(\hat{d}^\dagger_{\uparrow} \hat{d}_{\uparrow} + \hat{d}^\dagger_{\downarrow} \hat{d}_{\downarrow}) + U\hat{d}^\dagger_{\uparrow} \hat{d}_{\uparrow}\hat{d}^\dagger_{\downarrow} \hat{d}_{\downarrow}\,,\\
\hat{H}_{\alpha}&=\sum_{\boldsymbol{k},\sigma}
    \epsilon_{\alpha\boldsymbol{k}\sigma}
    \hat{c}^\dagger_{\alpha\boldsymbol{k}\sigma}\hat{c}_{\alpha\boldsymbol{k}\sigma} \quad(\text{for }\alpha \in\mathrm{L},\mathrm{R})\,,\\
\hat{V}_\alpha
&={\sum}_{\boldsymbol{k}, \sigma}  t_{\alpha\boldsymbol{k}\sigma} \hat{c}_{\alpha\boldsymbol{k}\sigma}^\dagger 
\hat{d}_{\sigma} 
 + \text{H.c.}\,,\\
\hat{H}_{\text{Z}}&=
    g_\mathrm{s}\frac{\mu_\mathrm{B} }{\hbar} 
    \boldsymbol{B}_\mathrm{ext} \cdot \boldsymbol{\hat{S}}\nonumber\\
    &=\frac{g_\mathrm{s}}{2} \mu_\mathrm{B}
    \boldsymbol{B}_\mathrm{ext} \cdot\sum_{\sigma,\sigma^\prime}\boldsymbol{\sigma}_{\sigma^\prime\sigma} \hat{d}^\dagger_{\sigma^\prime} \hat{d}_{\sigma}\,.
\end{align*}
The system Hamiltonian $\hat{H}_0$ describes the isolated \ac{QD} with a single energy level $\epsilon$ and on-site Coulomb repulsion $U$. 
The environment of the \ac{QD} consists of two fermionic baths corresponding to the left and right electrodes, described by the bath Hamiltonian $\hat{H}_{\mathrm{L,R}}$. Here, $\hat{c}_{\alpha\boldsymbol{k}\sigma}$ annihilates an electron with wavevector $\boldsymbol{k}$ and spin $\sigma$ in electrode $\alpha\in\lbrace L,R\rbrace$.
The interaction Hamiltonian $\hat{V}_\alpha$ describes the coupling of the \ac{QD} to electrode $\alpha$, given by the spin-dependent tunneling between the dot and the electrodes in form of a tight-binding Hamiltonian. Here, $t_{\alpha\boldsymbol{k}\sigma}$ is the tunneling amplitude for an electron of spin $\sigma$ to hop from the electrode to the dot. The spin index $\sigma\in\lbrace \uparrow,\downarrow\rbrace$ corresponds to the magnetic quantum number along the $z$-axis.
The Zeeman Hamiltonian $\hat{H}_{\text{Z}}$ describes the coupling of the dot spin to an external magnetic field $\boldsymbol{B}_\mathrm{ext}$, where $\boldsymbol{\sigma}\equiv \left( \sigma_x,\sigma_y,\sigma_z\right)$ are the Pauli matrices, $g_\mathrm{s}$ is the electron $g$-factor, and $\mu_\mathrm{B}$ is the Bohr magneton.

\begin{figure}[t!]
    \centering
    \includegraphics[scale=0.75]{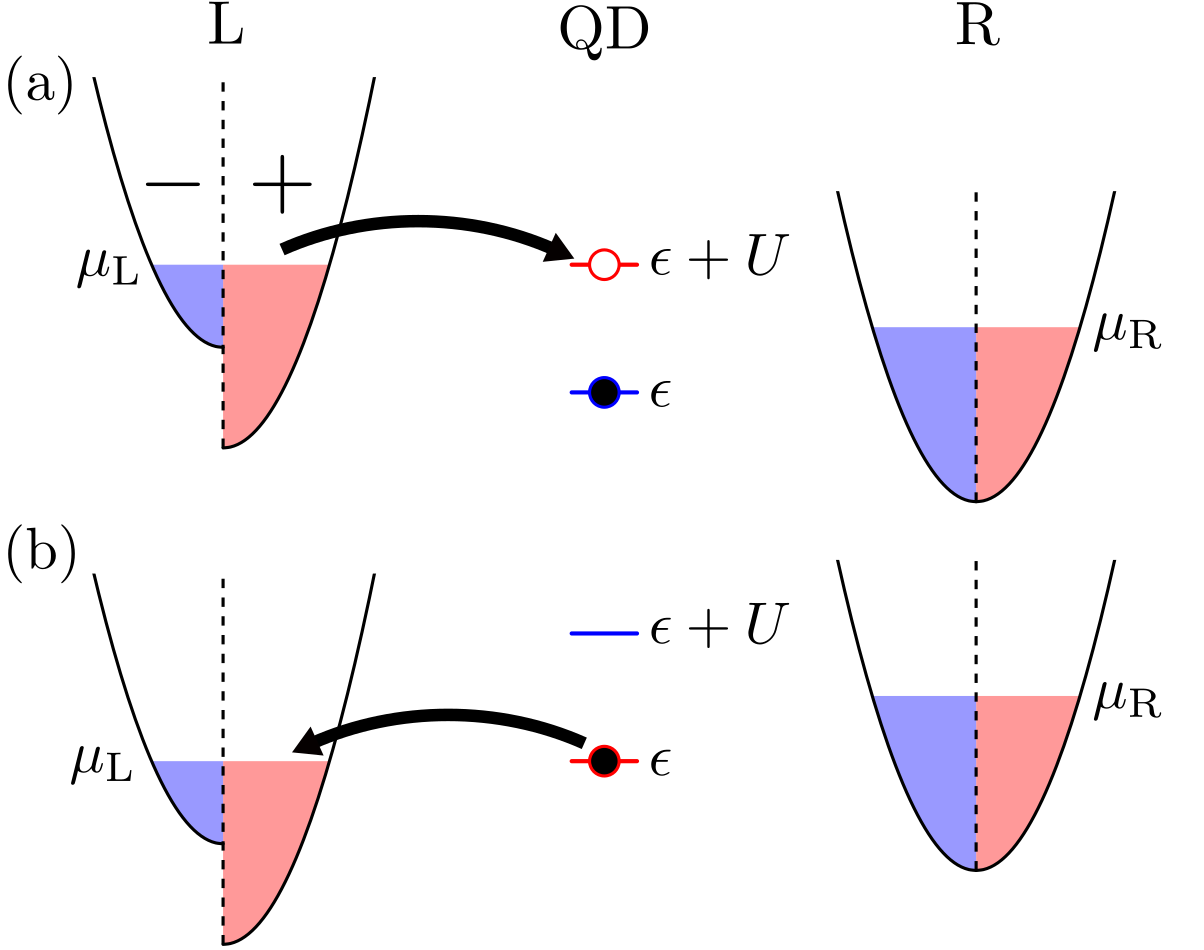}
    \caption{Schematic density of states diagrams showing spin-selective electron tunneling between a singly charged \ac{QD} and the left (L) and right (R) electrodes. 
    $+$ and $-$ denote majority and minority spin states in the magnetic electrode, which in general are not collinear with the $z$-quantization axis of the QD spin. $\epsilon$ and $\epsilon+U$ denote singly- and doubly-occupied QD states, respectively.
    (a) The \ac{QD} charges by hopping of an electron from the left (L) electrode with spin $+$, to form a singlet with an electron of opposite spin $-$. (b) At negative bias potential, the dot discharges by hopping of an electron onto the left electrode, occupying an empty electrode state with spin $+$.}
    \label{fig:sp-hopping}
\end{figure}

\subsection{Reduced 4x4 density matrix}\label{sec:4x4}
We briefly review the reduced density matrix approach in Fock-space, as developed in Ref.~\onlinecite{braun_theory_2004} to describe spin-dependent charge transport through a QD. In this formulation, lowest-order scattering diagrams are derived via a perturbative expansion of the Dyson series applied to the open quantum system governed by the Hamiltonian in Eq.~(\ref{eq:H}).
In addition to the coherent spin dynamics induced by the exchange field, we will show that incoherent tunneling between the electrodes and the \ac{QD} also contributes to spin evolution, driven by charge transfer processes.
We focus on a spin-1/2 \ac{QD} that hosts a single unpaired electron in the ground state. The dot can become doubly occupied (forming a singlet) under positive bias or emptied under negative bias. Since the empty and doubly occupied states are spinless, the relevant state space comprises four Fock states. Higher excited states are assumed to be energetically inaccessible within the applied voltage range.
In this approximation, the reduced density matrix is represented as a $4\times 4$ matrix in the basis of Eqs.~(\ref{eq:def_vac}-\ref{eq:def_two}):
\begin{equation}\label{eq:6}
\hat\rho \equiv \displaystyle \left(\begin{matrix}
\rho_{\emptyset\emptyset} &  &  & \\
 & \rho_{\uparrow\uparrow} & \rho_{\uparrow\downarrow} & \\
 & \rho_{\downarrow\uparrow} & \rho_{\downarrow\downarrow} & \\
 &  &  & \rho_{\mathrm{2}\mathrm{2}}
\end{matrix}\right)  .
\end{equation}

Off-diagonal elements coupling states with different charge numbers are neglected. First, coherences between charge sectors do not contribute to the instantaneous charge or spin expectation values. More importantly, strong electrode-induced dephasing and the separation of energies between spin and charge degrees of freedom ensure that off-diagonal terms can be assumed to be zero \cite{timm_tunneling_2008,mayrhofer_linear_2007}. Thus, neither the evolution of charge populations nor that of the spin state is subject to charge superpositions.

The time evolution of the reduced density matrix is governed by a generalized quantum master equation,
\begin{equation}\label{eq:me}
    \frac{\mathrm{d}}{\mathrm{d} t}\hat\rho(t)=
    -\frac{i}{\hbar}[  \hat{H}_0+\hat{H}_{\text{Z}},\hat\rho(t)] + \int\limits_{-\infty}^t \!\mathrm{d}t^\prime\, \hathat{\Sigma}(t^\prime,t) \hat\rho(t^\prime)\,.
\end{equation}
The right-hand side includes the unitary, coherent evolution under $\hat{H}_0+\hat{H}_{\text{Z}}$, as well as a coherent and a dissipative contribution from system-bath interactions encoded in the generalized self-energy tensor $\hathat{\Sigma}$ acting on Liouville space. 
In matrix form, the expression $\Sigma_{\chi^{\prime\prime}\chi^{\prime\prime\prime},\chi\chi^\prime}(t^\prime,t)\rho_{\chi\chi^\prime}(t^\prime)$ describes the probability rate for a transition from the state $|\chi\rangle\langle\chi^\prime|$ at time $t^\prime$ to the state $|\chi^{\prime\prime}\rangle\langle\chi^{\prime\prime\prime}|$ at time $t$, where $\chi, \chi^\prime$ and $ \chi^{\prime\prime},\chi^{\prime\prime\prime} \in\lbrace \emptyset,\uparrow,\downarrow\mathrm{2}\rbrace$.

\subsubsection{Lowest-order scattering diagrams}\label{sec:processes}
Assuming weak tunnel coupling, we retain only first-order terms in the perturbative expansion in the tunnel amplitudes, corresponding to transition rates $\Gamma_\alpha/\hbar\sim ||\hat{V}_\alpha||^2\hbar^{-2}$ in accordance with Fermi's golden rule. At this point, we emphasize that the system, bath, and interaction Hamiltonians leading to Eq.~(\ref{eq:me}) are assumed to be time-independent. Furthermore, in the Markovian limit, we may substitute $\rho(t^\prime)$ with $\rho(t)$, which is justified by considering short enough correlation times with the leads. Given the stationarity of the system-bath interactions this allows us to substitute the time-integral of the generalized self-energy with a time-independent tensor \cite{braun_theory_2004,konig_quantum_2005},
\begin{equation}
   \int\limits_{-\infty}^0 \!\mathrm{d}t^\prime\, \hathat{\Sigma}(t^\prime,t) \hat\rho(t^\prime)\to
   -\frac{i}{\hbar}
   \hathat{\Sigma}
   \hat\rho(t)\,.
\end{equation}

Two types of scattering processes emerge at this order: (i) real transitions involving electron tunneling into or out of the dot (charge transfer), and (ii) virtual fluctuations that affect the spin state without changing the dot's charge. The possible real tunneling processes between the spin-polarized electrode and the \ac{QD} are illustrated in Fig.~\ref{fig:sp-hopping}.

Figures~\ref{fig:scatter4} (a) and (b) show Keldysh diagrams representing real charge transitions in the notation of Ref.~\onlinecite{braun_theory_2004}, depicting the processes in which the \ac{QD} changes from a singly to a doubly charged state and back. 

\vspace{1em}
\input{diagrams/scatter1}

Figure~\ref{fig:scatter4} (c) and (d) show processes involving only virtual electron hopping that alter the dot's spin state while preserving its charge. The diagrams are directly related to the self-energy components $\Sigma_{\sigma^{\prime\prime}\sigma^{\prime\prime\prime},\sigma\sigma^\prime}$ (with $\sigma\in\lbrace \uparrow,\downarrow\rbrace$). For example

\input{diagrams/scatter-virt}

describes the transition  $|\!\uparrow\,\rangle\langle\,\uparrow\!| \overset{\mathrm{L}}{\longrightarrow}|\!\downarrow\,\rangle\langle\,\uparrow\!| $ due to lowest-order scattering with the left electrode, which is given by the sum of all possible processes. These processes correspond exclusively to virtual tunneling diagrams that only alter the spin state and result in an exchange field~\cite{braun_theory_2004}, effectively shifting the dot energy levels like a Zeeman term. This renormalization is Hermitian and conserves the norm $|\langle\hat{\boldsymbol{S}}\rangle|(t)$ at all times.

As an example, the change in the spin expectation value $\langle\hat{S}_y\rangle$ due to such a self-energy correction acting on the pure state  $\hat\rho(t_0)\equiv|\!\uparrow\,\rangle\langle\,\uparrow\!|$ is
\begin{align}
    \mathrm{d}\langle\hat{S}_y\rangle&=-i\frac{\hbar}{2}(\mathrm{d}\rho_{\uparrow\downarrow}-\mathrm{d}\rho_{\downarrow\uparrow})\nonumber\\    
    &=-i\frac{\hbar}{2}\sum_{\sigma,\sigma^\prime}-\frac{i}{\hbar}\mathrm{d}t\left(\Sigma_{\uparrow\downarrow,\sigma\sigma^\prime} - \Sigma_{\downarrow\uparrow,\sigma\sigma^\prime}\right)\rho_{\sigma\sigma^\prime}
    \nonumber\\
    &=-\frac{1}{2}\mathrm{d}t\left(\Sigma_{\uparrow\downarrow,\uparrow\uparrow} - \Sigma_{\downarrow\uparrow,\uparrow\uparrow}\right)\rho_{\uparrow\uparrow}\nonumber\\
    &=-\frac{1}{\hbar}\mathrm{d}t \left(\Sigma_{\uparrow\downarrow,\uparrow\uparrow}-\Sigma_{\downarrow\uparrow,\uparrow\uparrow}\right)\langle\hat{S}_z\rangle\nonumber\\
    &=\mathrm{d}t\, \frac{ g_\mathrm{s} \mu_\mathrm{B} }{\hbar} B_{\mathrm{xc},z} \langle\hat{S}_z\rangle
    =\mathrm{d}t\left(\langle\boldsymbol{S}\rangle \times \frac{ g_\mathrm{s} \mu_\mathrm{B} }{\hbar}\boldsymbol{B}_\mathrm{xc} \right)_y,
\end{align}
at $t=t_0$, where in this special caste the $x$-component $B_{\mathrm{xc},x}$ of the exchange field takes the simple form $-\left(\Sigma_{\uparrow\downarrow,\uparrow\uparrow}-\Sigma_{\downarrow\uparrow,\uparrow\uparrow}\right)/(g_\mathrm{s} \mu_\mathrm{B})$.
An expression for $\boldsymbol{B}_\mathrm{xc}$ in the general case is derived by a Schrieffer-Wolf transformation of terms $\hat{V}_\alpha$ in Eq.~(\ref{eq:H}), following Ref.~\onlinecite{konig_quantum_2005}, which leads to an effective Zeeman Hamiltonian $\hat{H}_\mathrm{xc}\propto 
    \boldsymbol{B}_{\mathrm{xc}} \cdot \boldsymbol{\hat{S}}$ with  $\boldsymbol{B}_{\mathrm{xc}}=\sum_\alpha\boldsymbol{B}_{\mathrm{xc},\alpha}$ consisting of field contributions from each electrode. 
    
\subsubsection{Quantum dot equations of motion}\label{sec:qd-eom}
A pivotal result of Ref.~\onlinecite{braun_theory_2004} is the derivation of the equations of motion for both charge and spin degrees of freedom in the reduced \ac{QD} system, as summarized in Appendix~\ref{app:eom}.
The time evolution of the charge-state populations is governed by Eq.~(\ref{eq:occuptations}), which describes the change in the diagonal elements of the density matrix corresponding to the uncharged state ($\rho_{\emptyset\emptyset}$), the singly occupied spin states ($\rho_{\uparrow\uparrow} + \rho_{\downarrow\downarrow}$), and the doubly charged singlet state ($\rho_{22}$). 

In parallel, the dynamics of the spin expectation value $\boldsymbol{S}$ is governed by Eq.~(\ref{eq:spins}). This equation comprises three distinct contributions: a DLT term $\left(\frac{\text{d}\boldsymbol{S}}{\mathrm{d}t}\right)_{\mathrm{D}}$ arising from spin-polarized charge transport between dot and the electrodes, a spin-relaxation term $\left(\frac{\text{d}\boldsymbol{S}}{\mathrm{d}t}\right)_{\mathrm{S}}$ reflecting decoherence and dissipation due to coupling with the reservoirs, and a precession term $\left( \frac{\mathrm{d}\boldsymbol{S}}{\mathrm{d}t} \right)_{\mathrm{F}} \propto \sum_{\alpha} \boldsymbol{S} \times  \boldsymbol{B}_{\mathrm{xc,\alpha}}$ that originates from the effective exchange field from each electrode. The latter can be interpreted as a torque generated by the dot's accumulated spin precessing around an effective magnetic field $\boldsymbol{B}_{\mathrm{xc,\alpha}}$ induced by spin-dependent virtual charge fluctuations. Note that Eq.~(\ref{eq:spins}) does not yet account for $\hat{H}_\mathrm{Z}$. The external field $\boldsymbol{B}_\mathrm{ext}$ is included by substituting $\sum_{\alpha} \boldsymbol{S} \times  \boldsymbol{B}_{\mathrm{xc,\alpha}} \rightarrow \boldsymbol{S} \times  \left(\, \sum_{\alpha}\! \boldsymbol{B}_{\mathrm{xc,\alpha}}+\boldsymbol{B}_\mathrm{ext}\right)$ into the field-like torque expression $\left( \frac{\mathrm{d}\boldsymbol{S}}{\mathrm{d}t} \right)_{\mathrm{F}}$.
Spin accumulation within the dot depends on the polarization of each electrode, denoted by $p_\alpha$, along the unit vector $\hat{n}_\alpha$ of each lead. Spin-selective charge transfer leads to DLT effects, which are discussed in more detail in Section~\ref{sec:DC-STT}. 

Overall, the generalized self-energy contributes to the dot's dynamics in two fundamental ways: First, it introduces a coherent renormalization of the system Hamiltonian, manifesting as an effective exchange field. Second, it accounts for dissipative effects, including changes in charge populations, spin accumulation, and spin-relaxation.
In the language of open quantum systems, the Hermitian part of this correction corresponds to the so-called Lamb shift, which is given by an additional Hamiltonian term $\hat{H}_\mathrm{xc}$. The non-Hermitian part gives rise to a dissipator, denoted by $\mathcal{D}$~\cite{manzano_short_2020}. Thus, the full reduced system dynamics can be written in the general form
\begin{equation}
    \frac{\mathrm{d}}{\mathrm{d}t}\hat\rho=
    -\frac{i}{\hbar}[\hat{H}_0 + \hat{H}_\mathrm{Z} + \hat{H}_\mathrm{xc},\hat\rho] + \mathcal{D}[\hat\rho]\,,
\end{equation}
where the exchange interaction shifts the energy levels of the dot according to
\begin{equation}
    \hat{H}_\mathrm{xc}= \sum_{\alpha=\mathrm{L},\mathrm{R}}\frac{ g_\mathrm{s} \mu_\mathrm{B} }{\hbar} 
    \boldsymbol{B}_{\mathrm{xc},\alpha} \cdot \boldsymbol{\hat{S}}\,.
\end{equation}
In our geometry only the contribution from the left electrode is present. Note that both the exchange contribution and the external magnetic field are expressed in formally equivalent terms by Zeeman couplings. 
The explicit form of the dissipator $\mathcal{D}[\hat\rho]$, which is consistent with the equations of motion for both charge and spin (Eqs.~(\ref{eq:occuptations})~and~(\ref{eq:spins})), is provided in the following section.

\subsection{Lindblad master equation}\label{sec:Lindblad-me}
To account for all lowest-order tunneling processes, we employ a master equation in which the time evolution of the reduced density matrix is linear in $\hat{\rho}$. We demonstrate that the equations of motion introduced in Ref.~\onlinecite{braun_theory_2004} can be reformulated in terms of a Lindblad-type master equation, where charge and discharge processes are represented as stochastic quantum jumps.
We begin by defining the unitary operator $\hat{U}(\theta,\varphi) \equiv \exp(i \theta\hat{S}_z/\hbar)\exp(i \varphi\hat{S}_y/\hbar)$
which maps the \ac{QD} states $|\!\uparrow\,\rangle$, $|\!\downarrow\,\rangle$ onto majority and minority spin states $|+\rangle$,$|-\rangle$ of the left electrode. The orientation of the left electrode's quantization axis $\mathbf{n}_L$ is parametrized by the Euler angles $\theta,\varphi$~\cite{Sakurai_Napolitano_2020}.
Under $\hat{U}(\theta,\varphi)$ the states introduced in Eqs.~(\ref{eq:def_vac})--(\ref{eq:def_two}), which span the reduced subspace of the QD, transform as
\begin{align}
    |\hspace{0.15em}\emptyset\hspace{0.15em}\rangle\longrightarrow& \hat{U}(\theta,\varphi)|\hspace{0.15em}\emptyset\hspace{0.15em}\rangle=|\hspace{0.15em}\emptyset\hspace{0.15em}\rangle\,,\\
    |\hspace{0.15em}2\hspace{0.15em}\rangle  \longrightarrow& \hat{U}(\theta,\varphi)|\hspace{0.15em}2\hspace{0.15em}\rangle=|\hspace{0.15em}2\hspace{0.15em}\rangle\,,\\
    |\!\uparrow\,\rangle\longrightarrow& \hat{U}(\theta,\varphi)|\!\uparrow\,\rangle\equiv|+\rangle\,,\\
    |\!\downarrow\,\rangle\longrightarrow& \hat{U}(\theta,\varphi)|\!\downarrow\,\rangle\equiv|-\rangle\,.
\end{align}
Note that the singlet states $|\hspace{0.15em}\emptyset\hspace{0.15em}\rangle$ and $|\hspace{0.15em}2\hspace{0.15em}\rangle$ are invariant under $\hat{U}(\theta,\varphi)$ and for the spin-$\tfrac{1}{2}$ the transformed states are given by
\begin{align}
    |+\rangle&= \;\;\;\cos(\theta/2)e^{-i\varphi/2}|\!\uparrow\,\rangle + \sin(\theta/2)e^{i\varphi/2}|\!\downarrow\,\rangle\,, \label{eq:def_pls}\\
    |-\rangle&= -\sin(\theta/2)e^{-i\varphi/2}|\!\uparrow\,\rangle + \cos(\theta/2)e^{i\varphi/2}|\!\downarrow\,\rangle\,.\label{eq:def_mns}
\end{align}

Using the above definitions for given values of $\theta$ and $\varphi$, we can introduce Lindblad operators associated with spin-dependent charging processes from the left electrode in the simple form
\begin{align}\label{eq:jumps1}
\hat{L}_1&\equiv |+\rangle \langle\hspace{0.15em}\emptyset\hspace{0.15em}|\,, \\
\label{eq:jumps2}
\hat{L}_2&\equiv|\hspace{0.15em}2\hspace{0.15em}\rangle \langle + |\,, \\ \label{eq:jumps3}
\hat{L}_3&\equiv|- \rangle \langle\hspace{0.15em}\emptyset\hspace{0.15em}|\,, \\ \label{eq:jumps4}
\hat{L}_4&\equiv|\hspace{0.15em}2\hspace{0.15em}\rangle \langle - |\,.
\end{align}
Their respective Hermitian adjoints correspond to the reverse (discharging) processes, which we define as $\hat{L}_n$ for labels $n=5,6,7,8$.
As an example, we can now associate the tunneling processes in Fig.~\ref{fig:sp-hopping}~(a)~and~(b) with Lindblad operators $\hat{L}_4=|\hspace{0.15em}2\hspace{0.15em}\rangle\langle-|$ and $\hat{L}_5\equiv\hat{L}^\dagger_1=|\hspace{0.15em}\emptyset\hspace{0.15em}\rangle\langle+|$, respectively.

The resulting equations of motion for the \ac{QD}, including the reduced Hamiltonian contributions, take the Lindblad form
\begin{align}\nonumber
\frac{\mathrm{d}}{\mathrm{d}t}\hat\rho(t) &=\ -\frac{i}{\hbar}
[\hat{H}_0+\hat{H}_{\mathrm{Z}}+\hat{H}_\mathrm{xc}, \hat\rho\left(t\right)]
\\
 &+\sum_{n,\alpha} \gamma_{n\alpha} 
\left(\hat{L}_n\hat\rho\left(t\right)\hat{L}_n^\dag - \frac{1}{2}\lbrace \hat{L}_n^\dag \hat{L}_n,\hat\rho\left(t\right)\rbrace
\right)
\,. \label{eq:lindblad-me}
\end{align}
The dissipative part of the equation includes a total of eight non-zero Lindblad operators $\hat{L}_{n}$, representing eight spin-dependent jump processes for each electrode. 
While the $+$,$-$ states are defined with respect to the spin-polarized left lead, we will use the same operators and jump operators for the unpolarized right lead, which has no preferential quantization axis.
Each process $n$ in electrode $\alpha$ occurs at a rate $\gamma_{n\alpha}$, which depends on the electron occupancy of the electrode, as described by the Fermi-Dirac distribution. 

Since the Lindblad operators enter twice into the dissipative part of the equation, we can pair each $\hat{L}_n$ to the square root of $\gamma_{n\alpha}$, for a given $\alpha$. This allows us to write the dissipator alternatively as $\sum_{n,\alpha} \hat{C}_{n\alpha}\hat\rho\hat{C}_{n\alpha}^\dag - \frac{1}{2}\lbrace \hat{C}_{n\alpha}^\dag \hat{C}_{n\alpha},\hat\rho\rbrace
$,
which is expressed in terms of collapse operators $\hat{C}_{n\alpha}\equiv\sqrt{\gamma_{n\alpha}} \hat{L}_n$ and take the explicit form
\begin{widetext}
\begin{align}\label{eq:cops12}
    &\sqrt{\gamma_{1\alpha}} \hat{L}_1\equiv\sqrt{\frac{  \Gamma_\alpha f^{\text{e}}_\alpha({\epsilon}) \left(1+p_\alpha\right)}{2\hbar}} |+ \rangle \langle\hspace{0.15em}\emptyset\hspace{0.15em}|\,,
    &\sqrt{\gamma_{2\alpha}} \hat{L}_2\equiv\sqrt{\frac{   \Gamma_\alpha f^{\text{e}}_\alpha({\epsilon+U}) \left(1-p_\alpha\right)}{2\hbar}}  |\hspace{0.15em}2\hspace{0.15em}\rangle \langle + |\,,\\\label{eq:cops34}
    &\sqrt{\gamma_{3\alpha}} \hat{L}_3\equiv\sqrt{\frac{  \Gamma_\alpha f^{\text{e}}_\alpha({\epsilon}) \left(1-p_\alpha\right)}{2\hbar}} |- \rangle \langle\hspace{0.15em}\emptyset\hspace{0.15em}|\,,
    &\sqrt{\gamma_{4\alpha}} \hat{L}_4\equiv\sqrt{\frac{   \Gamma_\alpha f^{\text{e}}_\alpha({\epsilon+U}) \left(1+p_\alpha\right)}{2\hbar}}  |\hspace{0.15em}2\hspace{0.15em}\rangle \langle - |\,,\\\label{eq:cops56}
    &\sqrt{\gamma_{5\alpha}} \hat{L}_5\equiv\sqrt{\frac{  \Gamma_\alpha f^{\text{h}}_\alpha({\epsilon}) \left(1+p_\alpha\right)}{2\hbar}} |\hspace{0.15em}\emptyset\hspace{0.15em}\rangle \langle +|\,,
    &\sqrt{\gamma_{6\alpha}} \hat{L}_6\equiv\sqrt{\frac{   \Gamma_\alpha f^{\text{h}}_\alpha({\epsilon+U}) \left(1 -p_\alpha\right)}{2\hbar}}  |+ \rangle \langle\hspace{0.15em}2\hspace{0.15em}|\,,\\\label{eq:cops78}
    &\sqrt{\gamma_{7\alpha}} \hat{L}_7\equiv\sqrt{\frac{  \Gamma_\alpha f^{\text{h}}_\alpha({\epsilon}) \left(1-p_\alpha\right)}{2\hbar}} |\hspace{0.15em}\emptyset\hspace{0.15em}\rangle \langle -|\,,
    &\sqrt{\gamma_{8\alpha}} \hat{L}_8\equiv\sqrt{\frac{   \Gamma_\alpha f^{\text{h}}_\alpha({\epsilon+U}) \left(1+p_\alpha\right)}{2\hbar}}  |- \rangle \langle\hspace{0.15em}2\hspace{0.15em}|\,.
\end{align}
\end{widetext}

Here, $f^{\text{e}}_\alpha(\epsilon)\equiv(e^{(\epsilon-\mu_{\alpha})/k_\mathrm{B}T}+1)^{-1}$ and $f^{\text{h}}_\alpha\equiv1-f^{\text{e}}_\alpha(\epsilon)$ denote the Fermi distribution for electrons ($\text{e}$) and holes ($\text{h}$) in electrode $\alpha$ at energy $\epsilon$, and electrochemical potential $\mu_\alpha$. A potential bias $-eV_{\mathrm{DC},\alpha}$ shifts $\mu_\alpha$ under application of a DC voltage on electrode $\alpha$. The voltage drop from left to right is then given by $V_\mathrm{DC}\equiv V_{\mathrm{DC},\mathrm{L}}-V_{\mathrm{DC},\mathrm{R}}$. 

We only consider the regime where the Zeeman splitting is small compared to the thermal energy $k_\mathrm{B} T$ and the Coulomb interaction $U$. 
In this limit, we neglect the energy shift due to the Zeeman energy and simply take the energy of the spin up and down level as $\epsilon$.
Consequently, the rates $\gamma_{n\alpha}$ become approximately independent of the spin splitting induced by either the external magnetic field $\boldsymbol{B}_\mathrm{ext}$ or the exchange field $\boldsymbol{B}_\mathrm{xc}$.

\subsection{Charge and spin currents}\label{sec:currents}
The master equation in Eq.~(\ref{eq:lindblad-me}) tracks the full temporal evolution of both populations and coherences in the charge and spin configurations of the QD. 
Because of charge conservation, any change in the dot occupation $-e\hat{N}$ must be balanced by a current to or from the electrodes. This is expressed as
\begin{equation}\label{eq:dQdt}
    -e\frac{\mathrm{d}}{\mathrm{d}t}\langle\hat{N}\rangle(t)=I_\mathrm{L}(t)+I_\mathrm{R}(t) \,,
\end{equation}
where the contribution of electrode $\alpha$ enters as a partial current $I_\alpha$ to the total current~\cite{koenig_quantum_1999}. The sign convention is such that $I_\alpha(t)>0$ describes electrons flowing from the QD to the electrode $\alpha$.
Note that the expectation value at time $t$ of an arbitrary operator $\hat{O}$, that commutes with $\hat{\rho}$ and is time-independent, evolves according to 
\begin{equation}    
\frac{\mathrm{d}}{\mathrm{d}t}\langle{\hat{O}}\rangle(t)=-\frac{i}{\hbar} \langle[ \hat{O},\hat{H}_\mathrm{eff}(t)]\rangle+\langle\mathcal{D^\dagger}[\hat{O}](t)\rangle\,,
\end{equation}
where all contributions to the unitary time-evolution of the reduced system are substituted into an effective Hamiltonian $\hat{H}_\mathrm{eff}$ \cite{breuer_heisenberg_1998,molmer_monte_1993}.
This allows for the charge and spin currents through the system to be expressed as quantum mechanical observables, specifically operators acting on the reduced Fock space.
The current operators at time $t$ are defined as follows for the charge,
\begin{align}
    \hat{I}_\alpha&=-e \sum_{n} \gamma_{n\alpha}\left( {\hat{L}^\dagger}_{n} \hat{N} {\hat{L}}_{n} - \frac{1}{2}\left\{{\hat{L}^\dagger}_{n}{\hat{L}}_{n},\hat{N} \right\}\right)\, ,\label{eq:current}
\end{align}
and spin currents,
\begin{align}
    \hat{\boldsymbol{I}}^\text{S}_\alpha&=-\frac{2e}{\hbar} \sum_{n} \gamma_{n\alpha}\left( {\hat{L}^\dagger}_{n} \hat{\boldsymbol{S}} {\hat{L}}_{n} - \frac{1}{2}\left\{{\hat{L}^\dagger}_{n}{\hat{L}}_{n},\hat{\boldsymbol{S}} \right\}\right)\, ,
\end{align}
The number operator $\hat{N}$ and the components of the spin vector operator $\boldsymbol{S} = (\hat{S}_x, \hat{S}_y, \hat{S}_z)$ are defined as
\begin{align}
    \hat{N}=\hat{N}_\uparrow + \hat{N}_\downarrow&=\hat{d}^\dagger_\uparrow \hat{d}_\uparrow + \hat{d}^\dagger_\downarrow \hat{d}_\downarrow\,,\\
    \hat{S_z} &= \frac{\hbar}{2}\left( \hat{N}_\uparrow - \hat{N}_\downarrow \right)\, ,\\
    \hat{S_x} &= \frac{\hbar}{2}\left( \hat{d}_\uparrow^\dag \hat{d}_\downarrow + \hat{d}_\downarrow^\dag \hat{d}_\uparrow \right)\,,\\
    \hat{S_y} &= \frac{\hbar}{2}\left( -i \hat{d}_\uparrow^\dag \hat{d}_\downarrow + i \hat{d}_\downarrow^\dag \hat{d}_\uparrow \right)\,.
\end{align}
Note that there are only dissipative contributions to the respective current operators, since both the spin and number operators commute with the reduced Hamiltonian, i.e., $[\hat{H}_\mathrm{eff},\hat{N}]=0$ and $[\hat{H}_\mathrm{eff},\hat{\boldsymbol{S}}\,]=\boldsymbol{0}$. 
Consequently, the charge and spin currents depend solely on terms involving the Lindblad operators $\hat{L}_n$. We now introduce quantities that represent the directed sequential charge and spin transport from electrode $\mathrm{L}$ to $\mathrm{R}$, namely
\begin{align}
    \hat{I} &\equiv \frac{1}{2}\left(\hat{I}_\mathrm{L} - \hat{I}_\mathrm{R}\right)\,,\\
    \hat{\boldsymbol{I}}^\text{S} &\equiv \frac{1}{2}\left(\hat{\boldsymbol{I}}^\text{S}_\mathrm{L} - \hat{\boldsymbol{I}}^\text{S}_\mathrm{R}\right)\,.
\end{align} Their expectation values $\langle \hat{I}\rangle (t)$ and $\langle \hat{\boldsymbol{I}}^\text{S}_\mathrm{R} \rangle (t)$ are time-local quantities that reflect the instantaneous probability of charge and spin transfer, respectively, given the instantaneous state $\rho(t)$ of the system.
Note that $\langle \hat{I}\rangle (t)$ and $\langle \hat{\boldsymbol{I}}^\text{S}_\mathrm{R} \rangle (t)$ are generally non-zero even if transport through only one electrode occurs. This reflects sequential transport even in the case of no through-current flowing from $\mathrm{L}$ to $\mathrm{R}$, and before a steady state is reached. The steady state cases defined by $\frac{\mathrm{d}}{\mathrm{d}t}\langle\hat{N}\rangle=0$ and $\frac{\mathrm{d}}{\mathrm{d}t}\langle\hat{\mathbf{S}}\rangle=0$, respectively, correspond to
\begin{align}
 \langle \hat{I}\rangle (t) =\langle \hat{I}_\mathrm{L}\rangle (t) = -\langle \hat{I}_\mathrm{R}\rangle (t)\,,\\
 \langle \hat{\boldsymbol{I}}^\mathrm{S}\rangle (t) =\langle \hat{\boldsymbol{I}}^\mathrm{S}_\mathrm{L}\rangle (t) = -\langle \hat{\boldsymbol{I}}^\mathrm{S}_\mathrm{R}\rangle (t)\,.
\end{align}
Hence, the introduced quantities $\langle \hat{I}\rangle (t)$ and $\langle \hat{\boldsymbol{I}}^\mathrm{S}\rangle (t) $ are equivalent to the respective physical charge and spin currents through the QD under steady state conditions.
The observable value of the current at time $t$ is obtained by taking the trace over the product of the current operator and the density matrix,
\begin{equation}
   I(t)\equiv \mathrm{tr}\left\{\hat{I}\hat{\rho}(t)\right\} \,.
\end{equation}

\subsection{Phenomenology of quantum spin torques}\label{sec:pheno}
\begin{figure}
    \centering
    \includegraphics[scale=0.75]{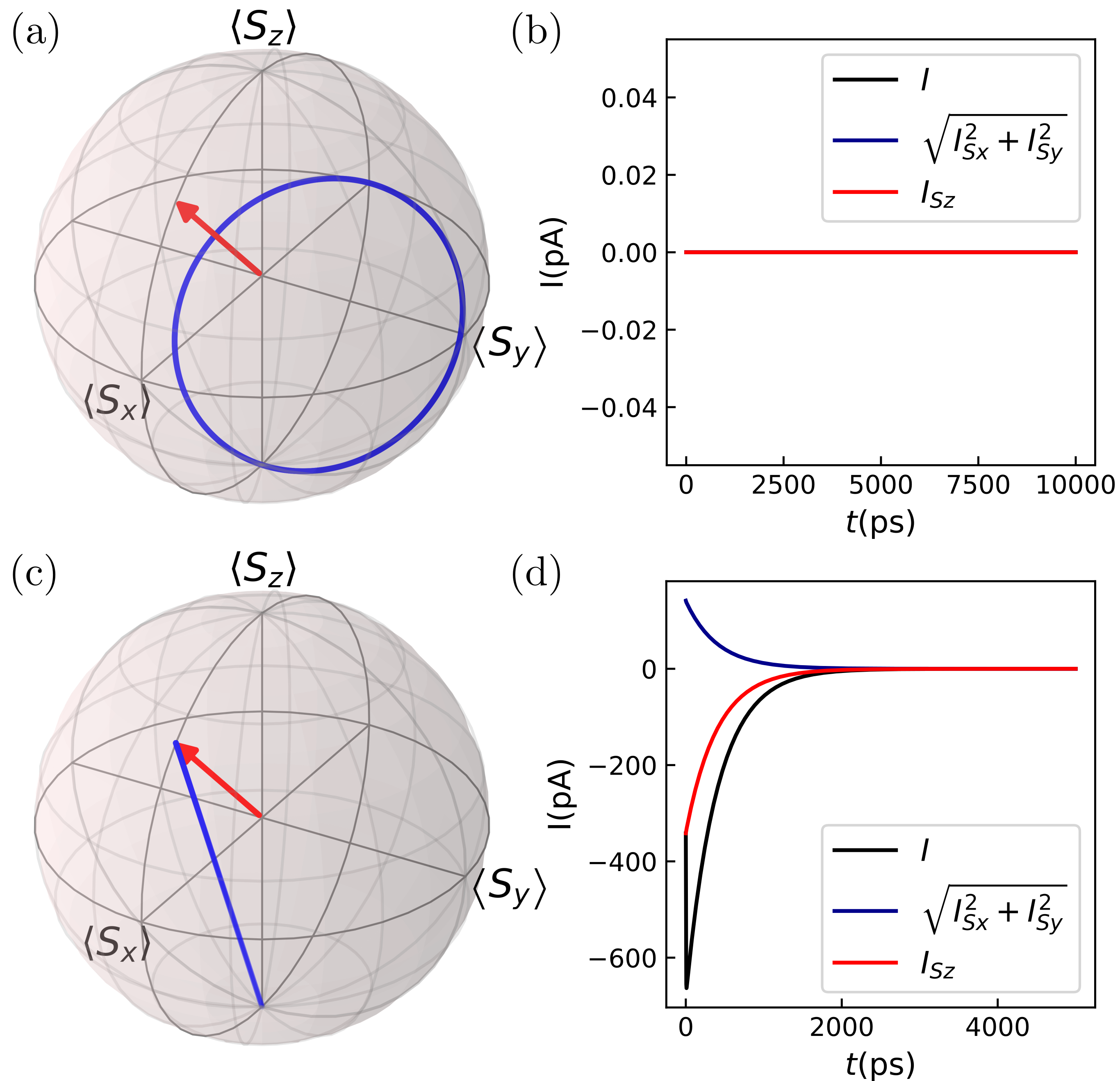}
\caption{Spin dynamics and tunneling current induced by FLT (a,b) and DLT (c,d) under a DC bias. (a) Larmor precession at zero external field, driven solely by a constant exchange field $\boldsymbol{B}_{\mathrm{xc},\mathrm{L}}$ parallel to unit vector $\boldsymbol{n}_\mathrm{L}$ indicated by the red arrow at angle $\theta=\pi/4$ to the $z$-axis. The blue trace represents the trajectory of the QD's spin starting from the state $|\!\downarrow\,\rangle$ ($\langle S_\mathrm{z} \rangle = -1$): the spin precesses in a plane orthogonal to $\boldsymbol{n}_\mathrm{L}$ (b) The net charge ($I$) and spin currents ($I_S$) are zero since only virtual hopping is present. 
  (c)  Spin rotation induced by the DLT due to current injection with rate $\gamma_{n\mathrm{L}}=0.005\hbar\, \mathrm{rad}\!\cdot\!\mathrm{ps}^{-1} \approx 0.0033\;\mathrm{meV}$. The spin state, initially in the $|\!\downarrow\,\rangle$ state, aligns with the left electrode's polarization along $\boldsymbol{n}_\mathrm{L}$. (d) Nonequilibrium charge and spin-currents dynamics due to the DLT. } 
    \label{fig:DC}
\end{figure}

Prior to discussing the effects of spin torques in QDs in more detail, we compare here the description of spin torques emerging from the QD's equations of motion and the phenomenological spin torque expressions adopted in spintronics.
In the configuration where only the left electrode is spin-polarized, the spin dynamics of the \ac{QD} can be described by the effective equation
\begin{equation}
    \frac{\mathrm{d}\boldsymbol{S}}{\mathrm{d}t}=
    \left(\frac{\mathrm{d}\boldsymbol{S}}{\mathrm{d}t}\right)_\mathrm{D}
    -
    \frac{\boldsymbol{S}}{\tau_\mathrm{S}}
    +
    \frac{g_\mathrm{s} \mu_\mathrm{B}}{\hbar}\boldsymbol{S}\times (\boldsymbol{B}_{\mathrm{xc},\mathrm{L}}+\boldsymbol{B}_\mathrm{ext})\,.\label{eq:Braun_spin_dyn}
\end{equation}
The first term on the right-hand side describes the DLT.
This term tends to align the spin of the \ac{QD} parallel (antiparallel) to the polarization direction of the left electrode, depending on whether the electrons flow from the left to the right (right to the left) electrode, which is controlled by the bias $V_\mathrm{DC}$.
The second term describes spin-relaxation due to tunnel coupling with the spin-polarized electrode. The relaxation time is given by  \cite{braun_theory_2004}

\begin{equation}
    \frac{1}{\tau_\mathrm{S}} = \sum_\alpha \frac{\Gamma_\alpha}{\hbar}\left(f_\alpha^\mathrm{h}(\epsilon)+f_\alpha^\mathrm{e}(\epsilon+U)\right)\,.\label{eq:spinlifetime}
\end{equation}
The third term is an FLT arising from an exchange-field $\boldsymbol{B}_{\mathrm{xc},\mathrm{L}}$ and an externally applied $\boldsymbol{B}_{\mathrm{ext}}$. The exchange field stems from interaction-induced virtual tunneling processes (see diagrammatic processes in Eq.~(\ref{eq:8})).

To connect this description to more general spintronic models, we compare Eq.~(\ref{eq:Braun_spin_dyn}) with the spin-diffusion equations, commonly used in the study of spin transport in magnetic multilayers and mesoscopic spintronic devices~\cite{shpiro_self-consistent_2003},
\begin{align}\label{eq:spintronics}
   \frac{\mathrm{d}\boldsymbol{M}_\mathrm{1}}{\mathrm{d}t}=&
   -\gamma_0a\boldsymbol{M}_\mathrm{1}\times\left(\boldsymbol{M}_\mathrm{2}\times\boldsymbol{M}_\mathrm{1}\right)
   +
   \alpha\boldsymbol{M}_\mathrm{1}\times\frac{\mathrm{d}\boldsymbol{M}_\mathrm{1}}{\mathrm{d}t}\nonumber\\
   &
   -\gamma_0\boldsymbol{M}_\mathrm{1}\times\left(\mathrm{H} + b\boldsymbol{M}_2\right)
   ,
\end{align}
where $\boldsymbol{M}_\mathrm{1}$ is the local magnetization of the free layer and $\boldsymbol{M}_2$ the pinned magnetization of the reference layer, which is separated by a nonmagnetic spacer from the free layer. These two magnetizations correspond to the QD and electrode spin polarizations, respectively. The Gilbert damping parameter is $\alpha$, the gyromagnetic ratio $\gamma_0$, and the parameters $a$ and $b$ characterize the damping-like and field-like torques, respectively, that the conduction electrons polarized by $\boldsymbol{M}_\mathrm{2}$ exert on $\boldsymbol{M}_\mathrm{1}$.

Each term in Eq.~(\ref{eq:spintronics}) has a qualitative counterpart in Eq.~(\ref{eq:Braun_spin_dyn}): the spin relaxation term $\sim\boldsymbol{M}_\mathrm{1}\times\frac{\mathrm{d}\boldsymbol{M}_\mathrm{1}}{\mathrm{d}t}$ parallels the quantum dot’s spin-relaxation; torque terms of the form $\sim\boldsymbol{M}_\mathrm{1}\times \boldsymbol{M}_2$ and $\sim\boldsymbol{M}_\mathrm{1}\times \boldsymbol{H}$ correspond to contributions of the exchange field $\boldsymbol{S} \times \boldsymbol{B}_{\mathrm{xc},\mathrm{L}}$ or the external magnetic field $\boldsymbol{S} \times \boldsymbol{B}_{\mathrm{ext}}$; and the damping-like torque term $\sim\boldsymbol{M}_\mathrm{1}\!\times\!\left(\boldsymbol{M}_2\!\times\!\boldsymbol{M}_1\right)$ is analogous to the spin accumulation $\left(\frac{\mathrm{d}\boldsymbol{S}}{\mathrm{d}t}\right)_\mathrm{D}$.
Notably, the exchange field in Eq.~(\ref{eq:Braun_spin_dyn}) contributes a purely unitary evolution of the spin dynamics, in contrast to the dissipative DLT and relaxation terms, which reduce the norm of the spin $|\langle\boldsymbol{S}\rangle|$ over time. This reflects the open-system character of the quantum dot, where decoherence arises due to charge transport. 

The approach in Ref.~\onlinecite{braun_theory_2004} offers a microscopic, quantum-consistent description of single-spin dynamics, allowing the coherent (Hamiltonian) contribution from the exchange field to be clearly separated from dissipative effects, such as spin accumulation and relaxation. In our Lindblad formalism (Eq.~(\ref{eq:lindblad-me})), this distinction naturally emerges: the coherent FLT enters via the renormalized system Hamiltonian, while damping-like dynamics appear in the dissipator, originating from quantum jump processes.

Thus, to arrive at a single-spin quantum interpretation of the damping-like spin torque, one must examine the effects of the dissipative mechanisms encoded in the Lindblad operators. In the following section, we explore the resulting spin dynamics through simulations, highlighting how these processes give rise to damping-like torques on the \ac{QD} spin.

\section{DC transport}\label{sec:DC-STT}
In this section, we examine steady-state transport phenomena in a molecular \ac{QD} coupled to spin-polarized electrodes. Several key effects arise in such spin-dependent tunneling setups: (i) spin accumulation, in which the dot spin becomes aligned with the polarization of the injecting electrode due to spin-selective tunneling; (ii) field-dependent spin blockade, where, for a dissipative torque, conductance minima occur as a magnetic field is swept through zero, reflecting suppression of spin transport due to coherent precession and relaxation mechanisms~\cite{braun_hanle_2005,busz_hanle_2023}; and (iii) spin blockade in systems of coupled spins, where spin selection rules hinder transport through spin-singlet and triplet configurations. Each of these phenomena reflects the interplay between quantum coherence, dissipation, and spin-selection rules inherent to the open quantum system. In the following subsections, we demonstrate how these effects naturally emerge within our charge-fluctuation model and characterize them through real-time simulations of the spin dynamics. The electrical current, computed as the expectation value of the corresponding Lindblad-derived current operator, serves as the central observable connecting theory to experiment.

To facilitate the discussion, we assume that the left electrode is fully spin polarized ($p_\mathrm{L}=1$) and the right electrode is nonmagnetic ($p_\mathrm{R}=0$). Furthermore, we align the spin-polarization $p_\mathrm{L}\!\cdot\!\boldsymbol{n}_\mathrm{L}$ of the left electrode (tip) with the $x$-axis ($\theta=0.5\,\pi$) if not stated otherwise. The rates are given by $\Gamma_\mathrm{L}/\hbar=0.005\, \mathrm{rad}\!\cdot\!\mathrm{ps}^{-1}$ and $\Gamma_\mathrm{R}/\hbar=0.5\, \mathrm{rad}\!\cdot\!\mathrm{ps}^{-1}$, which corresponds to $0.0033\;\mathrm{meV}$ and $0.3291\;\mathrm{meV}$ couplings. We assume the tip to keep a fixed polarization in the field sweep range without reversing with the field direction. 
The system Hamiltonian is defined by parameters corresponding to $\epsilon=-700\;\mathrm{meV}$, $U=1000\;\mathrm{meV}$ and the bias $V_\mathrm{DC}=400\;\mathrm{mV}$.

\subsection{DC spin torque on a single spin}\label{sec:DC-STT:ST}
Figure~\ref{fig:DC} compares the dynamics of a single spin subject to the FLT due to the exchange field and the DLT due to dissipative transport processes. The FLT leads to spin precession without current flow, as shown in panels (a)~and~(b). The effects of this torque have been widely discussed in the literature~\cite{koppens_driven_2006,nowack_coherent_2007,pla_single-atom_2012,yang_coherent_2019,lado_exchange_2017,willke_coherent_2021}. 

In the following, we focus on the effects of the DLT, as shown in panels (c)~and~(d).
In this scenario, a finite bias voltage enables sequential tunneling: an electron enters from the spin-polarized left electrode to doubly occupy the \ac{QD} and subsequently exits into the unpolarized right electrode. The resulting spin dynamics illustrates how the dot spin aligns with the polarization direction of the injecting electrode and how this alignment influences the steady-state current. An example of the time evolution of the dot's density matrix in this two-step sequential tunneling process is
\begin{equation}
    |- \rangle \langle - | 
    \overset{\mathrm{L}}{\longrightarrow} 
    |\hspace{0.15em}2\hspace{0.15em}\rangle \langle \hspace{0.15em}2\hspace{0.15em}|
    \overset{\mathrm{R}}{\longrightarrow} 
    \frac{1}{2}|+ \rangle \langle + | + \frac{1}{2}|- \rangle \langle - |\,.\label{eq:process-init}
\end{equation}
Here, we assume that the dot is initially in the $|- \rangle \langle - |$ state; the intermediate doubly occupied state is not virtual, but may have a short lifetime if $\Gamma_\mathrm{L}\ll\Gamma_\mathrm{R}$. Assuming that the left electrode is fully spin-polarized and the right electrode is nonmagnetic, Pauli exclusion permits tunneling only when the dot spin is anti-aligned with the left electrode's polarization. Consequently, the system evolves toward a steady state in which the dot's spin is aligned with the electrode polarization,
\begin{equation}
    |+ \rangle \langle + | 
    \longrightarrow
    |+ \rangle \langle + |\,.
\end{equation}
At this point, no further current flows since the transport is spin-blockaded.

This alignment mechanism remains valid for arbitrary orientations of the left electrode's polarization. 
For example, Fig.~\ref{fig:DC}(c) shows the case where the left electrode's polarization is tilted $45^{\circ}$ toward the $x$-axis relative to $z$. Starting from the state $|\!\downarrow\,\rangle$, the DLT drives the dot toward the aligned state $|+\rangle$.
Figure~\ref{fig:DC}(d) shows the corresponding real-time evolution of the charge and spin current components.

Notably, the presence of intermediate charge and spin fluctuations leads to a mixed character of the dot's state, implying that the \ac{QD} is not  strictly confined to a spin-1/2 Hilbert space. 
However, the steady state $|+\rangle\langle+|$ again represents a pure spin-1/2 configuration. The condition $\Gamma_\mathrm{L}\ll\Gamma_\mathrm{R}$ ensures that the average charge accumulation remains small. This illustrates how environmental interactions drive decoherence and relaxation in a single-spin system, deviating from purely unitary dynamics~\cite{braun_theory_2004, braun_hanle_2005}.
This process corresponds to the spin accumulation mechanism described in Ref.~\onlinecite{braun_theory_2004}, where the dot's spin aligns with the polarization of the spin-polarized electrode. It is inherently dissipative as angular momentum is transferred from the \ac{QD} to the reservoir. When only the left electrode is polarized, the spin relaxation time is approximately given by $\hbar/\Gamma_\mathrm{L}$. Further examples of DLT, caused by sequential charge transfer, and an effective $2\times2$ description of the spin evolution are given in Appendix~\ref{app:4x4_to_2x2}.

\subsection{DC Hanle effect for a single spin}\label{sec:DC-Hanle}
The Hanle effect describes the precession and dephasing of spin-polarized electrons in a nonmagnetic system under an applied transverse magnetic field~\cite{johnson_interfacial_1985}. It is a key experimental signature of spin transport and spin coherence, widely used in mesoscopic semiconductors, metals, and two-dimensional systems. The Hanle effect also offers a direct means of detecting spin accumulation induced by the DLT in \ac{QD}s by probing how external magnetic fields influence spin dynamics and current flow~\cite{braun_hanle_2005,mcmillan_image_2020,busz_hanle_2023,kovarik_spin_2024}. In such transport-based Hanle measurements, as the external field is swept from zero to a finite field, the conductance increases due to the precessional dephasing of spin-blockaded states.  The width and depth of the resulting conductance dip at zero field yield information about the spin lifetime and coherence of the system, accessible via DC currents~\cite{busz_hanle_2023}. 

\begin{figure}
    \centering
    \includegraphics[scale=0.75]{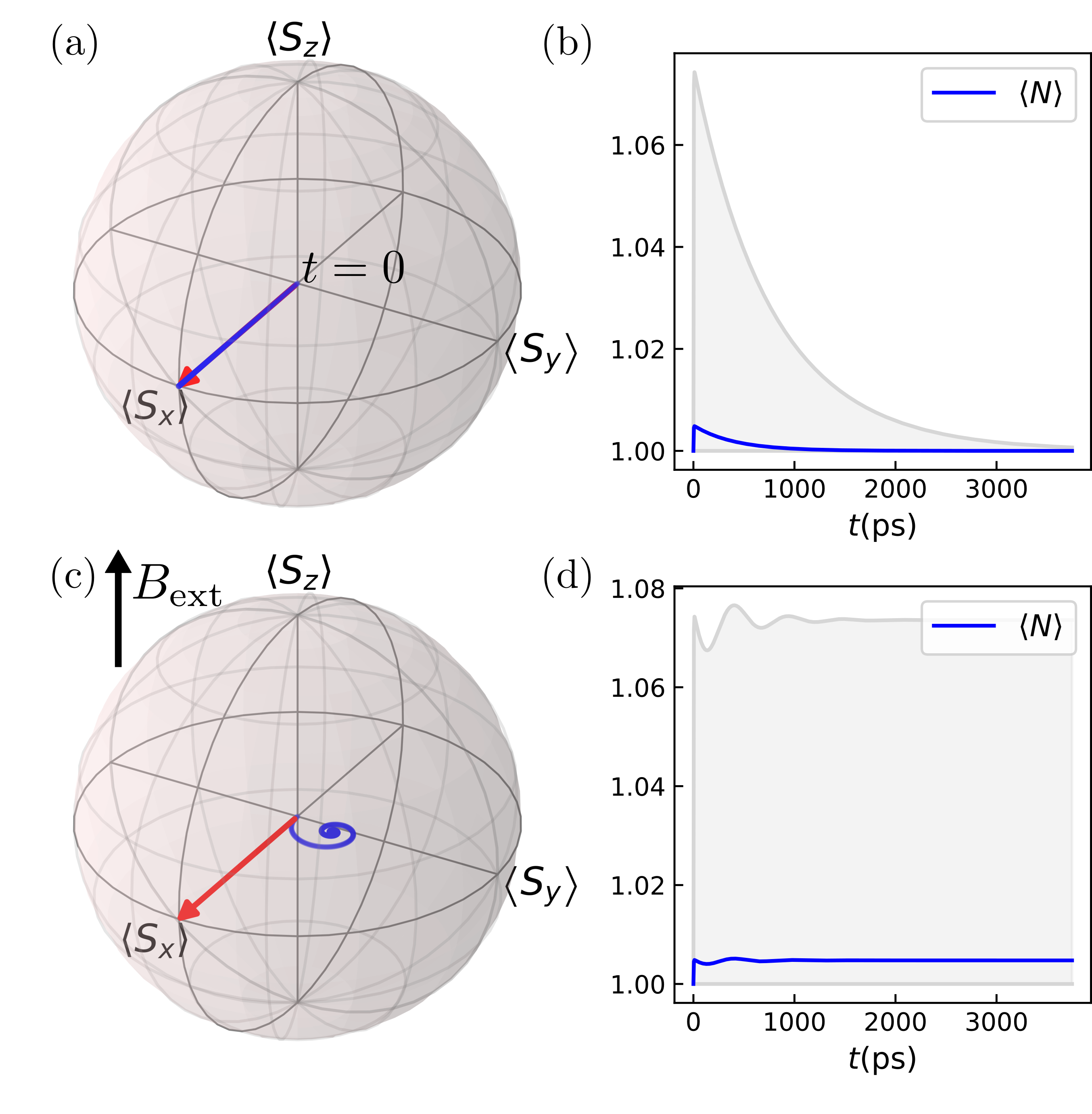}
    \caption{Spin and charge dynamics due to the DC Hanle effect. (a) In the absence of an external magnetic field, the injection of spin-polarized electrons from the left electrode leads to a complete Pauli spin blockade due to spin accumulation (blue line) along $\boldsymbol{n}_\mathrm{L}$ (red arrow). (b) This blockade prevents electron tunneling, thereby suppressing both charge accumulation beyond $\langle N\rangle =1$ and charge fluctuations. (c) Applying a finite field induces spin precession about the field axis (blue line), lifting the blockade. (d) As a result, continuous charge transport and accumulation $\langle N\rangle >1$ are enabled. The system evolves toward steady states, corresponding either to a blocked configuration with no net current (b) or a dynamic equilibrium between charging and draining (d). 
    Light gray intervals in (b) and (d) denote the statistical bounds of charge fluctuations, with a lower bound at $\langle \hat{N}\rangle=1$ and an upper bound given by one standard deviation: $\langle \hat{N}\rangle + \sqrt{\langle \hat{N}^2\rangle-\langle \hat{N}\rangle^2}$.}
    \label{fig:ZFFdynamics}
\end{figure}

\begin{figure}[!ht]
    \centering
    \includegraphics[scale=0.575]{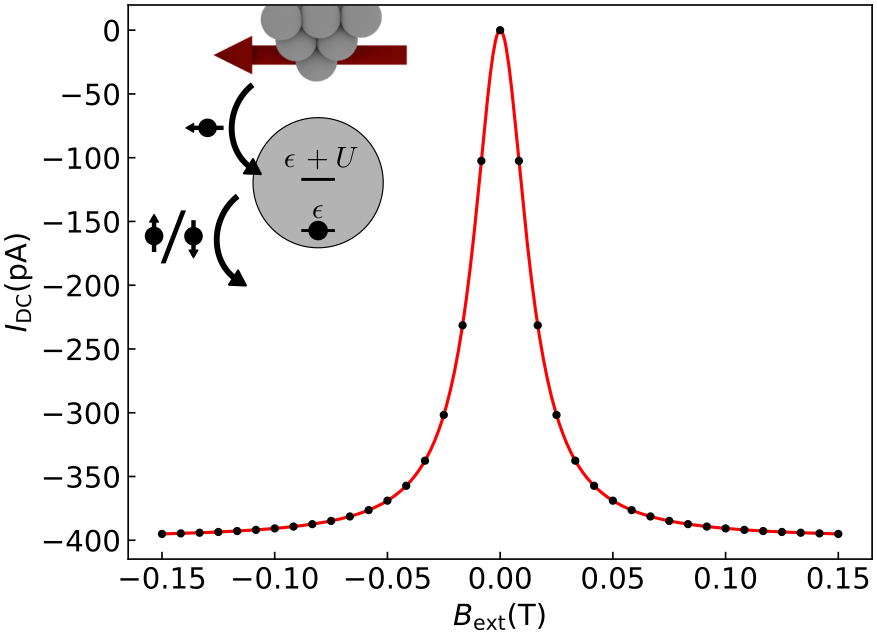}
    \caption{DC current-field characteristic of a molecular QD showing the Hanle effect.}
    \label{fig:zff-intro-no-coupling}
\end{figure}
In the sequential tunneling regime, starting from an initially unpolarized mixed state $\frac{1}{2}(|\!\uparrow\,\rangle\langle\,\uparrow\!| + |\!\downarrow\,\rangle\langle\,\downarrow\!|)$, the dot spin  $\langle\boldsymbol{S}\rangle(t)$ evolves toward alignment with the left electrode's polarization direction (taken to be along the $x$ axis, as shown in Fig.~\ref{fig:ZFFdynamics}(a)). This alignment, which is due to the DLT described in Sec.~\ref{sec:DC-STT:ST}, results in Pauli spin blockade, whereby Pauli exclusion and spin-conserving tunneling prevent the flow of electrons with the same spin as that occupying the dot~\cite{braun_hanle_2005,weinmann_spin_1995,ono_current_2002,koppens_driven_2006}. Consequently, the current vanishes and the dot charge stabilizes as $\langle\hat{N}\rangle\to 1$. Figure~\ref{fig:ZFFdynamics}(b) shows the corresponding real-time evolution of the dot occupation $\langle\hat{N}\rangle(t)$. Thus, at zero field, charge accumulation is hindered and a steady state is reached in which transport is effectively frozen due to spin-selective tunneling constraints.

When a finite magnetic field is applied, however, the dot spin undergoes Larmor precession, disrupting the full polarization along the electrode's axis and thereby lifting the spin blockade. This dynamics restores transport, as the system can now transition out of the blockaded state. Figure~\ref{fig:ZFFdynamics}(c,d) illustrates how spin precession and decoherence lead to a non-equilibrium state with sustained current and partial charge accumulation, $\langle \hat{N} \rangle > 1$.
At finite magnetic field, the \ac{QD} retains a spin polarization, but with reduced amplitude and rotated relative to the full polarization at zero field. As the field increases, the polarization amplitude decreases toward zero, while the spin orientation tilts further within the precession plane compared to the zero-field steady state.

The resulting field-dependent conductance profile, which is a hallmark of the Hanle effect, is shown in Fig.~\ref{fig:zff-intro-no-coupling}. The dip near zero field follows a Lorentzian function, where the full width at half maximum (FWHM) corresponds to the inverse spin lifetime $\tau_\mathrm{S}^{-1}$. According to Eq.~(\ref{eq:spinlifetime}) and in the limit $\Gamma_\mathrm{L}\ll\Gamma_R$, this corresponds to

\begin{equation}
    \frac{1}{\tau_\mathrm{S}} \approx \frac{\Gamma_\mathrm{L}}{\hbar} \left(f^{\text{h}}_\mathrm{L}(\epsilon)+f^{\text{e}}_\mathrm{R}(\epsilon + U)\right)\,.
\end{equation}
Comparison with the expression in Eq.~(\ref{eq:current}) and recognizing that charge accumulation vanishes in the limit $\Gamma_\mathrm{L}\ll\Gamma_R$, the current depends linearly only on $\Gamma_\mathrm{L}$. In particular, in the sequential tunneling regime with $f^\mathrm{h}_\mathrm{R}(\epsilon)\approx 0$, $f^\mathrm{e}_\mathrm{R}(\epsilon+U)\approx0$, and assuming no spin accumulation, both the left- and right-going currents $\langle\hat{I}_\mathrm{L}\rangle\propto \Gamma_\mathrm{L}f^\mathrm{h}(\epsilon)$ or $\langle\hat{I}_\mathrm{R}\rangle\propto \Gamma_\mathrm{L}f^\mathrm{e}(\epsilon+U)$ are limited by $\Gamma_\mathrm{L}$. Thus, both the current baseline (far from zero-field) and the linewidth depend linearly on $\Gamma_\mathrm{L}$ (while the linear regime holds for $\Gamma_\mathrm{L}\ll\Gamma_R$).

\subsection{Sensor-spectator dynamics}\label{sec:DC-sensor-spectator}
\subsubsection{Incoherent coupled spin dynamics}

\begin{figure}[!ht]
    \centering
    \includegraphics[scale=0.575]{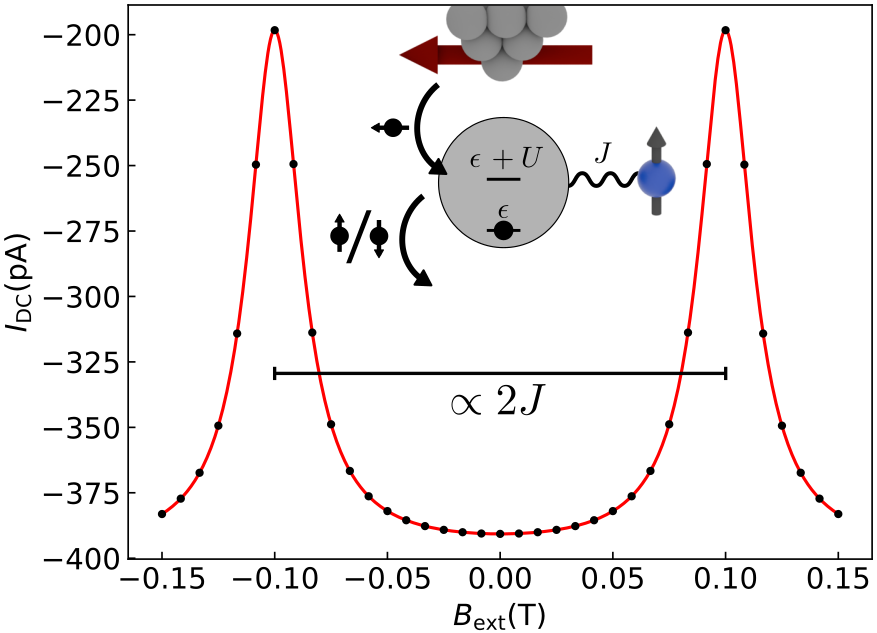}
    \caption{DC current-field characteristic of a molecular \ac{QD} coupled to an Ising spin.}
    \label{fig:zff-intro-coupling}
\end{figure}

We now examine the situation in which a transport-active \ac{QD} (\textit{sensor}) is exchange-coupled to a second localized spin (\textit{spectator}), as shown in the inset of Fig.~\ref{fig:zff-intro-coupling}~\cite{mcmillan_image_2020}. The spectator is assumed to be in a fully mixed state, with equal probabilities for spin-up $|\!\uparrow\,\rangle$ and spin-down $|\!\downarrow\,\rangle$. This spin does not participate directly in transport but couples to the sensor spin via an Ising-type interaction Hamiltonian,
 \begin{align}
     \hat{H}_{J} =
     &-2J/\hbar \left(\hat{S}_z\otimes\mathbb{1}\right)\left(\mathbb{1}\otimes\hat{\sigma}_z\right)
     = \nonumber\\
     &-J\left(|\!\uparrow\Uparrow\rangle\langle\uparrow\Uparrow\!| +|\!\downarrow\Downarrow\rangle\langle\downarrow\Downarrow\!|\right)\nonumber\\&+J\left(|\!\uparrow\Downarrow\rangle\langle\uparrow\Downarrow\!|+|\!\downarrow\Uparrow\rangle\langle\downarrow\Uparrow\!|\right)\,,
 \end{align}
where $J$ is the coupling constant between sensor and spectator. 
For simplicity, we assume that the mixed state of the spectator $\frac{1}{2}|\!\Uparrow\,\rangle \langle\,\Uparrow\!| + \frac{1}{2}|\!\Downarrow\,\rangle \langle\,\Downarrow\!|$ is not affected by the applied external field. Field-dependent occupation of the spectator states can be implemented in a straightforward way when necessary.

We consider the Hanle process described in Sec.~\ref{sec:DC-Hanle} for a blockaded sensor spin.
The influence of sensor-spectator coupling manifests itself as a splitting of the Hanle dip into two distinct minima, as shown in Fig.~\ref{fig:zff-intro-coupling}. Each dip corresponds to a different effective magnetic field determined by the spectator being in either the $|\!\uparrow\,\rangle$ or $|\!\downarrow\,\rangle$ state. Because the sensor spin experiences the combined effect of the external and exchange fields, the conductance minimum splits, reflecting the statistical mixture of internal fields. The relative height of the two dips encodes the occupation probabilities of the spectator spin states, and thus provides a sensitive probe of the local magnetic environment of the QD.

Such a configuration could enable direct electrical readout of atomic-scale classical or quantum bit states , as well as spin thermometry applications~\cite{del_castillo_theory_2025, natterer_reading_2017}. Unlike previous methods that require radio-frequency control, the spin-torque-induced readout presented here is passive and non-invasive. 
Resolving differences in Hanle dip strength due to unequal occupations of the spectator's spin states requires either stabilizing the left electrode (tip) polarization against reversal by the external field~\cite{kubetzka_spin-polarized_2002}, or stabilizing the spectator's magnetic population, e.g., through additional local exchange interactions.

\subsubsection{Coherent coupled spin dynamics}\label{sec:dimer-collapse}

\begin{figure*}
    \centering
    \includegraphics[scale=0.575]{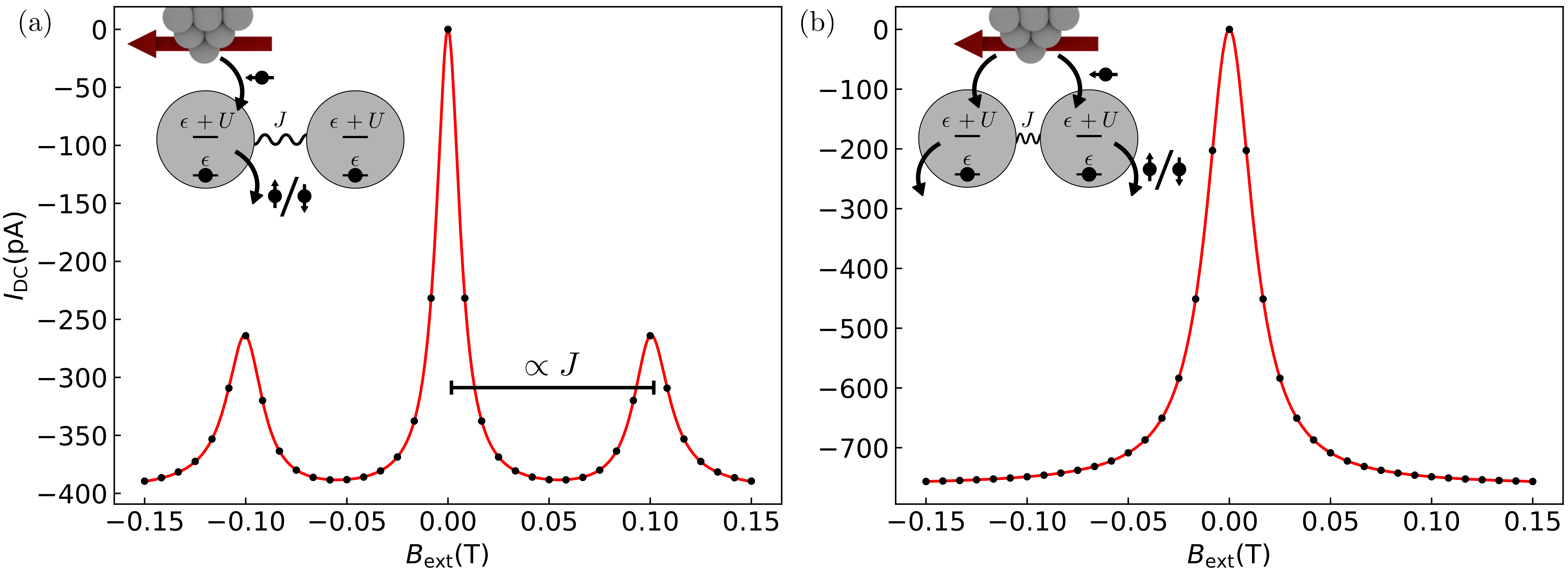}
    \caption{DC Hanle effect in coupled spin systems. (a) Magnetic field sweep near zero-field where spin-polarized transport occurs only through a \ac{QD} sensor coupled to a non-transported spectator spin. (b) Simultaneous spin-torque on both spins leads to disappearance of Hanle sidebands. Electron transport can occur either through the left or the right QD. Half of the total current, corresponding to transport through the left \ac{QD} only, is tracked for comparability with (a).}
    \label{fig:zff-collapse}
\end{figure*}

A richer scenario emerges when both the sensor and spectator are treated as coherent spin-1/2 systems coupled via a Heisenberg-type interaction,
 \begin{equation}\label{eq:Ham-J}
     \hat{H}_{J}=-\frac{2J}{\hbar^2} \left(\hat{\boldsymbol{S}}\otimes\mathbb{1}\right) \cdot \left(\mathbb{1}\otimes\hat{\boldsymbol{S}}\right)\,,
 \end{equation}
including an additional Zeeman term for the spectator spin.
This configuration was studied theoretically in Ref.~\onlinecite{mcmillan_image_2020}, revealing a characteristic three-peak structure in the current-field curve: a central dip at zero field and two symmetric sidebands at $B_\mathrm{ext} = \pm 2J/(g_\mathrm{s}\mu_\mathrm{B})$, as shown in Fig.~\ref{fig:zff-collapse}(a). The chosen sign convention in Eq.~(\ref{eq:Ham-J}) corresponds to a ferromagnetic, triplet ground state for $J>0$  when no external field is applied. The sign of $J$ is selective to whether the state $|\!\uparrow\uparrow\rangle$ or $|\!\downarrow\downarrow\rangle$ is mixed with the singlet $\left(|\!\uparrow\downarrow\rangle-|\!\downarrow\uparrow\rangle\right)/\sqrt{2}$ when $B_\mathrm{ext}$ is tuned to the respective sideband \cite{mcmillan_image_2020}. In this respect, the roles of the left and right sidebands are exchanged by inverting the sign of $J$.

The central dip arises from the familiar spin blockade at zero field, where the sensor aligns with the spin-polarized electrode and transport halts. When considering the sensor spin state only, its steady state $|+\rangle\langle+|$ is reached. For the combined system, the density matrix approaches the pure state $|++\rangle\langle++|$, which corresponds to a degenerate eigenstate since both spins are aligned and no external field is applied.
However, the sidebands reflect coherent dynamics involving both spins: At these fields, the effective magnetic field on the sensor vanishes due to cancellation by the exchange interaction, enabling formation of the blockaded states, which arise from a suppressed probability of finding the sensor spin in a $|-\rangle$ state. Formally, we compute the expectation value of $\left\langle\vphantom{\frac{}{}}|-\rangle\langle-|\otimes\mathbb{1}\right\rangle$, which is a measure of the hopping rate from the left electrode. By comparison with Eq.~(\ref{eq:current}), an expression for the current is obtained by the identity
\begin{equation}
    {\hat{L}^\dagger} \hat{N} {\hat{L}} - \frac{1}{2}\left\{{\hat{L}^\dagger}{\hat{L}},\hat{N} \right\} = |-\rangle\langle-|\otimes\mathbb{1}\,,
\end{equation}
where the number operator $\hat{N}$ and the Lindblad operator $\hat{L}\equiv\hat{L}_4\otimes\mathbb{1}=|\hspace{0.15em}2\hspace{0.15em}\rangle\langle-|\otimes\mathbb{1}$ are extended to act on the two-\ac{QD} Hilbert space. For long enough evolution times, the system's density matrix evolves to a mixed dynamical steady state, with only partial blocking of the current, which explains the reduced depth of the spectral side bands by $\approx1/3$ relative to the zero-field line, in accordance with Ref.~\onlinecite{mcmillan_image_2020}.
%

 We now consider the case that transport can occur through both dots. The extension to two QDs is implemented by defining collapse operators for both dots, $\hat{C}^\mathrm{A}_{n\alpha}\equiv\sqrt{\gamma^\mathrm{A}_{n\alpha}}\hat{L}_{n_\alpha}\otimes\mathbb{1}$ and $\hat{C}^\mathrm{B}_{n\alpha}\equiv\sqrt{\gamma^\mathrm{B}_{n\alpha}}\mathbb{1}\otimes\hat{L}_{n_\alpha}$, respectively. Here, $\gamma^i_{n\alpha}$ are the rates for dot $i\in\{\mathrm{A},\mathrm{B}\}$. The individual currents for dot $i$ are then computed according to Sec.~\ref{sec:currents}. The total (spin) current is the sum of the (spin) currents through both dot A and B.

When tunneling occurs through both the sensor and the spectator equally, as illustrated in the inset of Fig.~\ref{fig:zff-collapse}, the situation changes dramatically. A simultaneous DLT applied to both spins acts as a projective measurement in the spin-polarized electrode basis, collapsing the spectator's state required for the sidebands.
As a result, the spin coherence needed to observe a Pauli blockade at the sidebands is lost, and only the central dip remains (see Fig.~\ref{fig:zff-collapse}(b)). This phenomenon highlights the destructive role of stochastic transport pathways.

\section{AC transport}\label{sec:AC-STT}
While the Lindblad master equation [Eq.~(\ref{eq:lindblad-me})] effectively describes DLT and spin-initialization processes~\cite{kim_anisotropic_2022}, exciting EPR requires time-dependent driving of the dot's spin. In STM experiments, this is achieved by adding an RF voltage to the bias. In our model, this enters through a generalization to time-dependent rates $\gamma_{n\alpha} (t)$ and a time-dependent exchange Hamiltonian $\hat{H}_\mathrm{xc}(t)$.

Note that the promotion of $\gamma_{n\alpha}$ and $\hat{H}_\mathrm{xc}$ to time-dependent quantities is phenomenological and that a time-dependence of the electrode parameters directly contradicts the assumptions in Sec.~\ref{sec:processes}.
The diagrammatic framework reviewed in Sec.~\ref{sec:processes} applies only to the time-independent (DC) case. A generalization to a time-dependent driven \ac{QD} is not straightforward. Yet, a formal microscopic description for periodic driving using a Floquet master equation formalism is provided in Ref.~\onlinecite{reina-galvez_study_2021}, which modifies the hopping term in the interaction Hamiltonian in Eq.~(\ref{eq:6}) by time-modulated parameters $t_{\alpha\boldsymbol{k}\sigma}\rightarrow t_{\alpha}(t)$ \cite{reina-galvez_many-body_2023}. 

For a purely phenomenological approach with effective rates, however, one does not need to consider all microscopic processes. This may be desirable considering the
 complexity of the physical problem, which renders two difficulties in attempting a microscopic description: 
\begin{enumerate}
    \item A perturbative treatment of scattering may not be possible under a strong RF drive.
    \item A realistic model of a driven \ac{QD} should, in general, consider not only a modulation of the system but also of the environment and tunnel-coupling.
\end{enumerate} 
Different approaches in the literature consider either a modulation of the QD, or the tunneling barrier and/or the electrodes \cite{jauho_time-dependent_1994,andergassen_charge_2010,cavaliere_nonadiabatic_2009,shavit_generalized_2019,winkler_theory_2013,reina-galvez_study_2021}, which goes beyond the diagrammatic description in Sec.~\ref{sec:processes}. The ideal case presented in Sec.~\ref{sec:processes} without AC driving allows for explicit first-principle derivation of the time-independent rates $\gamma_{n\alpha}$ in Eq.~(\ref{eq:cops12})--(\ref{eq:cops78}) in the lowest tunneling order. Furthermore, realistic \ac{QD} systems are embedded in a highly complex environment, including, for example, phonon-assisted tunneling \cite{gawronski_imaging_2008,baxevanis_dynamics_2014} and coupling to vibrational modes of the system \cite{krane_vibrational_2025,eickhoff_inelastic_2020}. 

The master equation approach used here and Ref.~\onlinecite{mcmillan_image_2020} provides an effective description of experimental transport phenomena~\cite{kovarik_spin_2024}, without requiring explicit calculations of all microscopic processes and relaxation channels from first principles. Within this framework, $\hat{H}_\mathrm{xc}(t)$ and $\gamma_{n\alpha} (t)$ serve as model parameters that enable the interpretation of experimental observations, provided that the dynamics remains Markovian, ensuring the validity of a time-dependent Lindblad-type master equation~\cite{rivas_quantum_2012}. Thus, while the time-dependent parameters are introduced in an ad-hoc manner, their existence is justified as long as the dynamics are those of a memory-less evolution. A master equation describing Markov dynamics can be casted into the Lindblad form \cite{rivas_quantum_2012}. Hence, when considering negligible charge coherences, following the arguments in Sec~\ref{sec:4x4}, our Lindblad description with only eight time-dependent collapse operators per lead $\hat{C}_{n\alpha}(t)$ is generally valid. The influence of these parameters  was analyzed in Sec.~\ref{sec:DC-STT} and Sec.~\ref{sec:AC-STT} for a DC- and AC-driven \acp{QD}, respectively.

\subsection{AC-FLT driving}
The regime relevant for coherent spin dynamics is the Coulomb blockade regime~\cite{weymann_tunnel_2005, reina_galvez_cotunneling_2019, kovarik_spin_2024}, where sequential tunneling is exponentially suppressed and higher-order tunneling (e.g., co-tunneling) is dominant. Examples of second-order diagrams corresponding to cotunneling are shown in Appendix~\ref{app:cotunneling}. These second-order processes can induce Kondo-like exchange interactions~\cite{ternes_spin_2015, reina_galvez_cotunneling_2019, ast_theory_2024, schrieffer_relation_1966}. 

In the absence of sequential tunneling, we can describe the pure FLT contribution via a time-dependent effective Hamiltonian acting on the spin subspace
\begin{equation} \label{eq:Hxc}
    \hat{H}_\mathrm{xc}(t)=\sum_\alpha \frac{g_\mathrm{s}\mu_\mathrm{B}}{\hbar} \boldsymbol{\hat{S}} \cdot \left( \boldsymbol{B}_{\mathrm{xc},\alpha}^{(0)} + \boldsymbol{B}_{\mathrm{xc},\alpha}^{(1)}\cos(\omega t) \right)\,.
\end{equation}
In contrast to the DC-FLT featured in Fig.~\ref{fig:DC}, oscillating fields resulting from AC transport yield terms of the form $\boldsymbol{B}_{\mathrm{xc},\alpha}^{(1)}\cos(\omega t)$. The total AC contribution of the field is henceforth identified with an AC-FLT, also known as $B_1$-driving in the context of magnetic resonance or quantum information processing.

\begin{figure}
    \centering
    \includegraphics[scale=0.75]{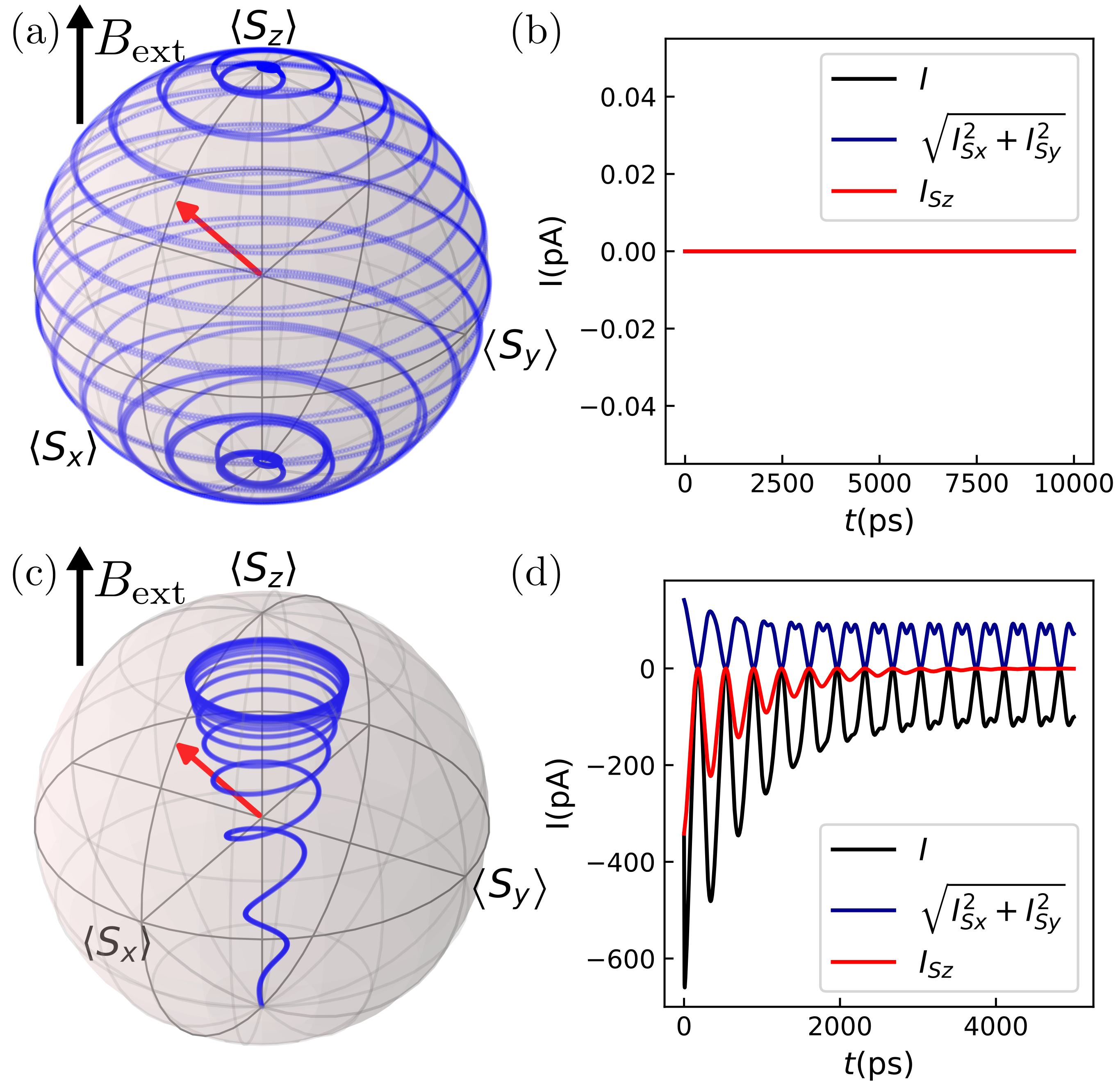}
    \caption{Spin dynamics and tunneling current induced by an AC-FLT and AC-DLT at resonance $\omega=\omega_0$ in an external field $B_\mathrm{ext}=\hbar \omega_0 / (g_\mathrm{s}\mu_\mathrm{B}) = 0.1\;\mathrm{T}$. (a) Rabi oscillations induced by the exchange field  $\boldsymbol{B}_{\mathrm{xc},\mathrm{L}}^{(1)} \cos (\omega t)$. $\boldsymbol{B}_{\mathrm{xc},\mathrm{L}}^{(1)}$ is oriented at $\theta=\pi/4$ along $\boldsymbol{n}_\mathrm{L}$ in the x-z plane (red arrow). (b) The net charge and spin current are zero in the pure AC-FLT  case. 
  (c) EPR driven by an AC-DLT with modulation of the rates $ \gamma_{n\mathrm{L}} (t)=\gamma^{(1)}_{n\mathrm{L}}\cos^2(\omega t/2)$ assuming sequential tunneling from a left electrode with full spin polarization along $\theta=\pi/4$ (red arrow). The rates relevant for charging and discharging are $\gamma^{(1)}_{4\mathrm{L}}=0.005\hbar\, \mathrm{rad}\!\cdot\!\mathrm{ps}^{-1} \approx 0.0033\;\mathrm{meV}$, and $\gamma_{6\mathrm{R}}(t)=\gamma_{8\mathrm{R}}(t)=0.5\, \mathrm{rad}\!\cdot\!\mathrm{ps}^{-1} \approx 0.33\;\mathrm{meV}/\hbar$, respectively. The QD spin, initially in the $|\!\downarrow\,\rangle$ state, evolves into a steady precessional mixed state about the external field. (d) Non-equilibrium charge and spin-currents due to periodically-modulated electron transfer corresponding to a pure AC-DLT. } 
    \label{fig:AC}
\end{figure}

A special case arises when we assume $\boldsymbol{B}_{\mathrm{xc}, \mathrm{R}}^{(k)} = 0$ for $k=0,1$, such that only the left electrode contributes to the exchange field. For a generic geometry where $\boldsymbol{B}_\mathrm{ext}$ is not collinear with $\boldsymbol{B}_{\mathrm{xc},\mathrm{L}}^{(1)}$, resonance occurs when the driving frequency matches the Larmor frequency $\omega_0={g_\mathrm{s}\mu_\mathrm{B}  |\boldsymbol{B}_\mathrm{ext} + \boldsymbol{B}_{\mathrm{xc}, \mathrm{L}}^{(0)}|}/\hbar$~\cite{lado_exchange_2017}. This results in Rabi oscillations at frequency $\Omega=g_\mathrm{s}\mu_\mathrm{B}|\boldsymbol{B}_{\mathrm{xc}, \mathrm{L}}^{(1)}| \cos(\tilde\theta)/\hbar$, where $\tilde\theta$ is the angle between the total static field $\boldsymbol{B}_\mathrm{ext} + \boldsymbol{B}_{\mathrm{xc}, \mathrm{L}}^{(0)}$ and the oscillating component $\boldsymbol{B}_{\mathrm{xc}, \mathrm{L}}^{(1)}$. Figures~\ref{fig:AC}(a) and (b) illustrate the AC-FLT dynamics induced by the oscillatory exchange field.

\subsection{AC-DLT driving and AC Hanle effect}\label{sec:AC-STT-QD}
In this section, we consider physical scenarios where sequential transport is allowed and dominates over FLT to the point that it can be neglected (with exclusion of the external field $\boldsymbol{B}_\mathrm{ext}$).
Here we demonstrate the effect of substituting $\gamma_{n\alpha}$ defined in Eqs.~(\ref{eq:cops12})--(\ref{eq:cops78}) by generally time-dependent effective rates $\gamma_{n\alpha}(t)$ leading to a time-dependent damping of the QD.

\begin{figure}
    \centering
    \includegraphics[scale=0.75]{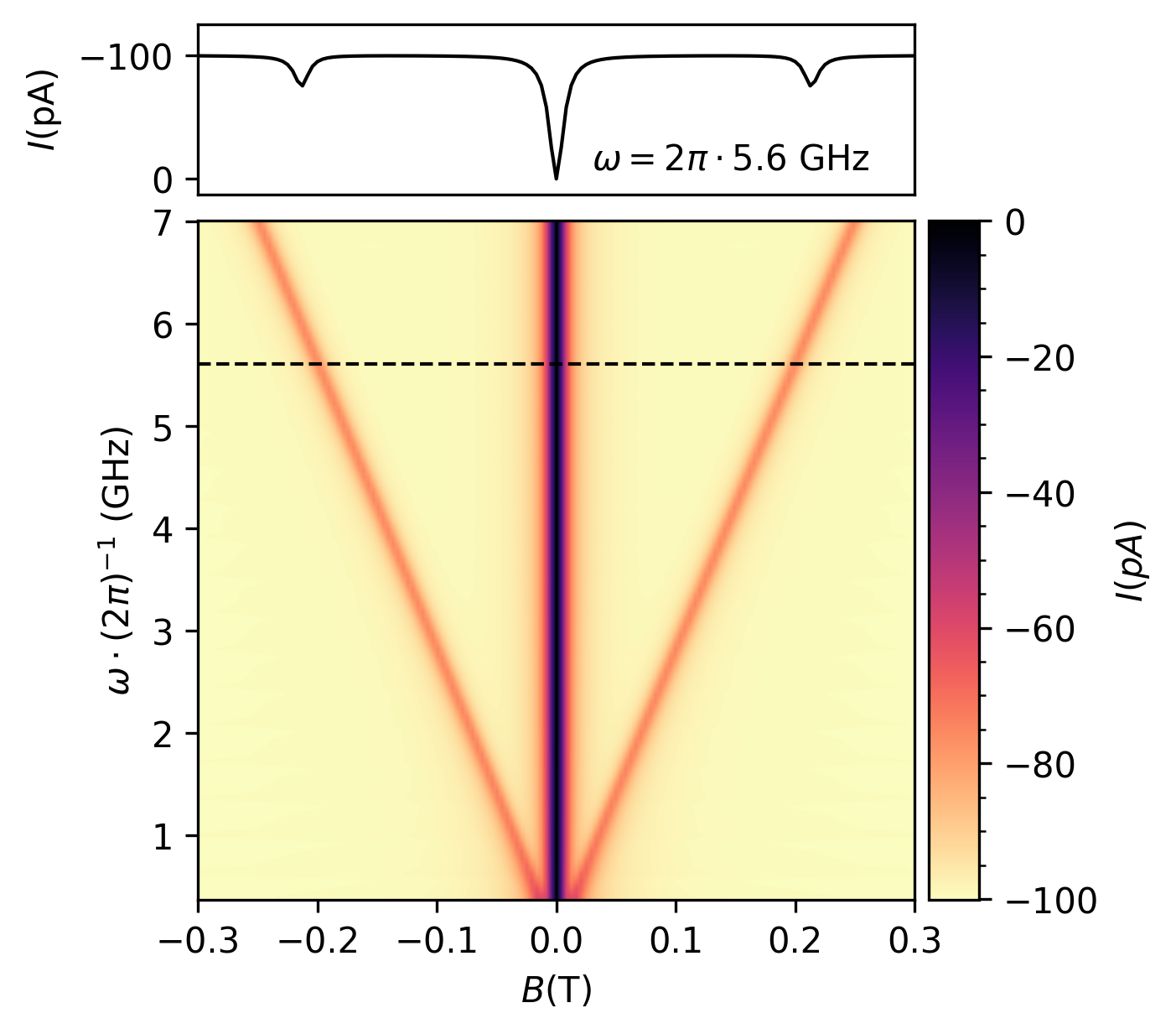}
    \caption{Emergence of DC and AC Hanle effect by periodically modulated DLT-driving at varying driving frequencies $\omega\cdot(2\pi)^{-1}$.
    The finite transport rates are
    $\gamma_{4\mathrm{L}}^{(1)}= 0.005\, \mathrm{rad}\!\cdot\!\mathrm{ps}^{-1}$ and $\gamma_{6\mathrm{R}}^{(0)}, \gamma_{8\mathrm{R}}^{(0)} = 0.5\, \mathrm{rad}\!\cdot\!\mathrm{ps}^{-1} \approx 0.33\;\mathrm{meV}/\hbar$.}
    \label{fig:map-freq}
\end{figure}

This AC-DLT, in particular for a periodic modulation of the stochastic transport rates, can lead to the driving of magnetic resonance of the QD when $\hbar\omega$ matches the transition energies of the spin system \cite{kovarik_spin_2024,reina-galvez_contrasting_2025}. 
In order to implement both DC- and AC-DLT we consider finite time-dependent Lindblad rates of the form,
\begin{equation}\label{eq:gamma}
    \gamma_{n\alpha} (t) = \gamma_{n\alpha}^{(0)} + \gamma^{(1)}_{n\alpha}\frac{1+\cos(\omega t)}{2} \,.
\end{equation}
Figure~\ref{fig:map-freq} shows simulated time-averaged current spectra, exhibiting spectral lines that correspond to pure DC- and AC-DLT (assuming no FLT) for a dynamical steady state. The spectral lines resemble experimentally observable resonances in related QD setups \cite{koppens_driven_2006,kovarik_spin_2024}. The AC case exhibits the characteristic of magnetic resonance driving, associated with diverting spectral lines, depending on the external field and the driving rate $\omega$. The DC-Hanle resonance remains centered around $B_\mathrm{ext}=0$ as the line position is independent of $\omega$.

To understand the underlying dynamics for AC-DLT driven EPR, we first need to determine how a steady state is reached.
In resonantly driven molecular \acp{QD}, a precessing steady state is reached within a characteristic lifetime $T_{\mathrm{D}}$, which has been identified as the timescale in which the spin follows the oscillating RF electric field~\cite{kovarik_spin_2024}, and is directly linked to the left electrode coupling as described by $\gamma_{n\mathrm{L}}(t)$ in our QD setup ~\cite{reina-galvez_study_2021,mcmillan_image_2020}. Since in the lab frame and at finite B-fields, any finite transversal spin polarization, is oscillatory, we consider the rotating frame.
At a finite field, we describe the rotating frame spin states as 
\begin{equation} 
|\pm\rangle_\mathrm{rot} = e^{-i\hat{S}_z \omega_0 t/\hbar} |\pm\rangle\,, 
\end{equation} 
which leads to the density matrix
\begin{align}
&\left\{|\pm \rangle \langle \pm |\right\}_\mathrm{rot}=
\frac{1}{2} \hat\sigma_0 \pm \frac{1}2\cos(\theta)\hat{\sigma}_z \nonumber\\
&\pm \frac{1}{2}\sin(\theta)\left(e^{-i\varphi- i\omega_0 t}|\uparrow\,\rangle\langle\,\downarrow| + e^{+i\varphi+ i\omega_0 t}|\downarrow\,\rangle\langle\,\uparrow|  \right)\label{eq:coherences}\\
&=
\frac{1}{2} \hat\sigma_0 \pm \frac{1}{2}\cos(\theta)\hat\sigma_z\nonumber\\
&\pm \frac{1}{2}\sin(\theta)\left(\cos(\varphi+\omega_0 t))\hat\sigma_x + \sin(\varphi+\omega_0 t)\hat\sigma_y  \right)\,,
\end{align}
where $\hat\sigma_0\equiv|\!\uparrow\,\rangle\langle\,\uparrow\!|+|\!\downarrow\,\rangle\langle\,\downarrow\!|$.
In general, the expectation values with respect to the lab-frame states $|\pm\rangle$ are given by the projections
\begin{align}
    \left\langle\vphantom{\frac{}{}}\left\{|\pm \rangle \langle \pm |\right\}_\mathrm{rot}\right\rangle_\pm
    &=
    \cos^2(\theta)+\sin^2(\theta)\cos^2 \left(\frac{1}{2}\omega_0 t\right)\,,\label{eq:rotproj}\\
   \left\langle\vphantom{\frac{}{}}\left\{|\mp \rangle \langle \mp |\right\}_\mathrm{rot}\right\rangle_\pm
    &=
   \sin^2(\theta)\sin^2 \left(\frac{1}{2}\omega_0 t\right)\,.
\end{align}

Since only the coherences $|\!\uparrow\,\rangle\langle\,\downarrow\!|$ and $|\!\downarrow\,\rangle\langle\,\uparrow\!|$ have time-dependent prefactors in Eq.~(\ref{eq:coherences}), it is sufficient to consider only the precessing contributions to the density matrix proportional to $\cos(\varphi+\omega_0 t))\hat\sigma_x + \sin(\varphi+\omega_0 t)\hat\sigma_y$. 
Without loss of generality, we consider $\varphi=0$ and $\theta=\pi/2$, which corresponds to a spin confined in the xy-plane. This leads to simple expressions for the time-dependent projections
\begin{align}
    \left\langle\vphantom{\frac{}{}}\left\{|\pm \rangle \langle \pm |\right\}_\mathrm{rot}\right\rangle_\pm
    &=
    \cos^2 \left(\frac{1}{2}\omega_0 t\right)
    =\frac{1+\cos(\omega_0 t)}{2}\,,\label{eq:rotproj++}\\
   \left\langle\vphantom{\frac{}{}}\left\{|\mp \rangle \langle \mp |\right\}_\mathrm{rot}\right\rangle_\pm
    &=
   \sin^2 \left(\frac{1}{2}\omega_0 t\right)\
   =\frac{1-\cos(\omega_0 t)}{2}\,.
\end{align}
These projections reflect the precessing spin state components along the x-axis as is evident from the identity
\begin{align}
\nonumber \\ \label{eq:precessing_states}
    &\frac{1}{2}\hat\sigma_0\pm\cos(\omega_0t)\hat{\sigma}_x
   =\nonumber\\
   &\cos^2\left(\frac{1}{2}\omega_0 t\right)|\pm \rangle \langle \pm | 
   + \sin^2\left(\frac{1}{2}\omega_0 t\right)|\mp \rangle \langle \mp | 
   \,.
\end{align}
In analogy to DC spin initialization, see Eq.~(\ref{eq:process-init}), the spin dynamics can be modeled through sequential charging and discharging for the precessing system,
\begin{align}
    &\left\{|- \rangle \langle - |\right\}_\mathrm{rot}
    \overset{\mathrm{L}}{\longrightarrow} 
    |\hspace{0.15em}2\hspace{0.15em}\rangle \langle\hspace{0.15em}\mathrm{2}\hspace{0.15em}|
    \overset{\mathrm{R}}{\longrightarrow}
    \frac{1}{2}\left\{|+ \rangle \langle + |\right\}_\mathrm{rot}
    + 
    \frac{1}{2}\left\{|- \rangle \langle - |\right\}_\mathrm{rot}\,.\label{eq:process-init-Larmor}
\end{align}
This process causes dynamical spin relaxation which physically depends on the timescale of the coupling, and is here described by $\hbar/{\gamma^{(1)}_{n\mathrm{L}}}$.
A time-independent dissipator results in equal decay of both $|\pm \rangle \langle \pm |_\mathrm{rot}$ components; however, a time-dependent dissipator can yield dynamical selection~\cite{shakirov_spin_2019, kovarik_spin_2024, reina-galvez_study_2021, gunderson_floquet_2021}. 
Specifically, modulating the dissipator with $\gamma_{n\mathrm{L}}(t) =\gamma^{(1)}_{n\mathrm{L}}\cos^2(\omega t/2)$ and tuning $\omega = \omega_0$ leads to state-selective relaxation. 
As shown in Fig.~\ref{fig:AC} (c) and (d), this modulation suppresses the decay of the $\left\{|+ \rangle \langle + |\right\}_\mathrm{rot}$ component and enhances the decay of $\left\{|- \rangle \langle - |\right\}_\mathrm{rot}$, effectively acting as a homodyne filter. This process can be understood as a phase-locking mechanism on the time scale of $T_\mathrm{D}\sim\hbar/\gamma^{(1)}_{n\mathrm{L}}$ and is equivalent to an AC-DLT-driven \ac{EPR}~\cite{kovarik_spin_2024}, analogous to the DC-DLT generated by spin accumulation described in Sec.~\ref{sec:DC-STT}. In analogy to the DC Hanle effect (Sec.~\ref{sec:DC-STT}), the AC-DLT can be viewed as an \textit{AC Hanle effect}~\cite{bazaliy_alternating_2024}.

In both cases, spin accumulation generates net spin currents and buildup of blockade, as a result of the spin-polarized electrode, and the detuning from resonance with the external field results in the observed Hanle effects. 

\begin{figure}
    \centering
    \includegraphics[scale=0.75]{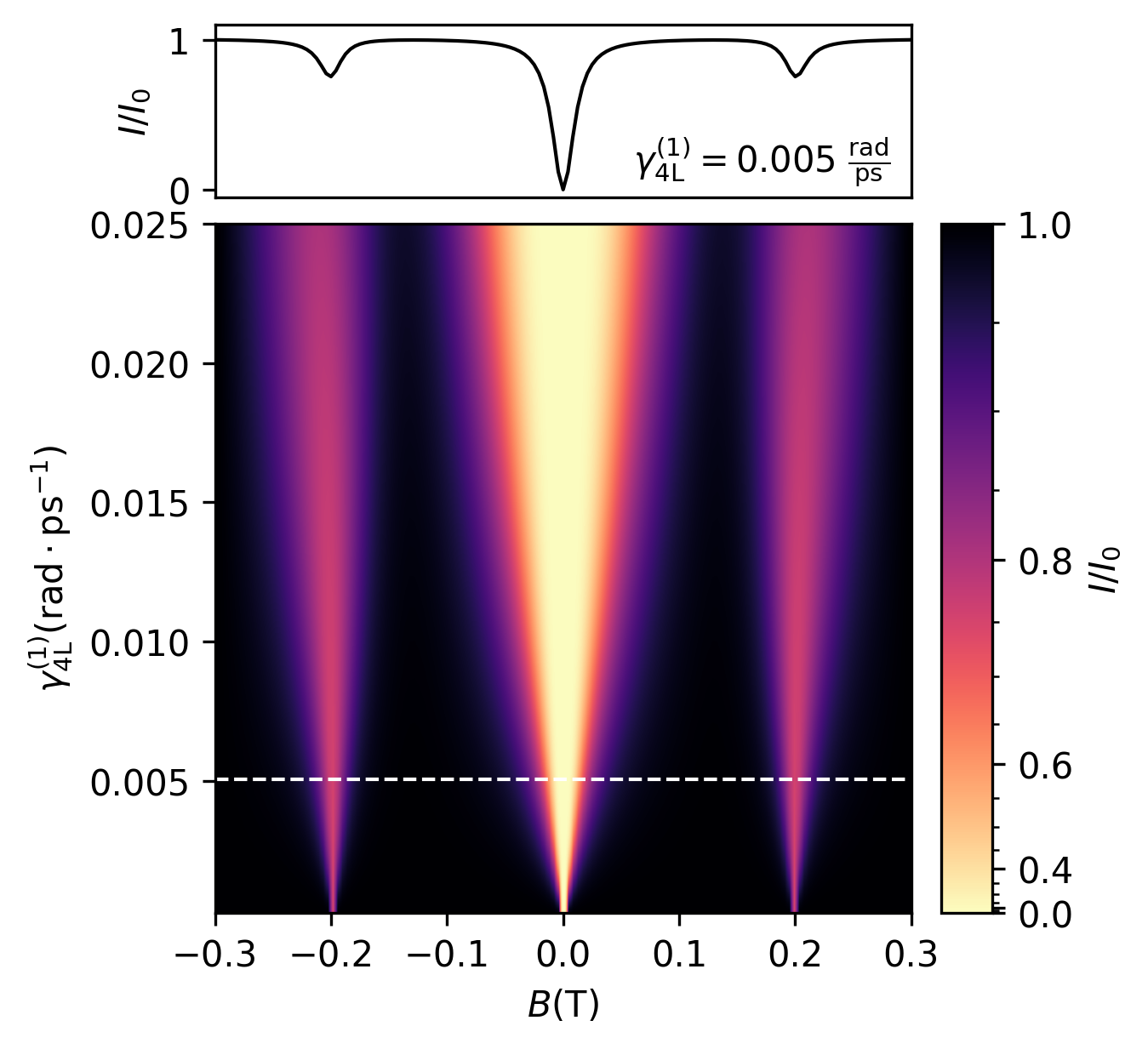}
    \caption{Broadening of DC and AC Hanle linewidths with increasing AC-DLT driving strength at $\theta=\pi/4$. The singly charged dot charges with a rate $\gamma_{4\mathrm{L}}^{(1)}\cos^2(\omega t)$. The discharging of the doubly occupied dot is fixed with $\gamma_{6\mathrm{R}}^{(0)}, \gamma_{8\mathrm{R}}^{(0)} = 0.5\, \mathrm{rad}\!\cdot\!\mathrm{ps}^{-1} \approx 0.33\;\mathrm{meV}/\hbar$. The AC Hanle features at $\pm0.2$,T, correspond to where the Larmor rate $\omega_0$ matches the driving rate $\omega$ and are equivalent to DLT-driven EPR.}
    \label{fig:mapgamma}
\end{figure}

We now focus on the effect the scale of $\gamma^{(1)}_{n\mathrm{L}}$ has on the spectrum under DLT-driving.
Figure~\ref{fig:mapgamma} shows the tunneling current as a function of ${B}_\mathrm{ext}$ for different values of the driving amplitude described by $\gamma_{n\mathrm{L}}^{(1)}$. Increasing $\gamma^{(1)}_\mathrm{nL}$ broadens both DC and AC Hanle features. The DC contribution simply arises from the 0th harmonic contribution of the $\cos^2(\omega t/2)$ modulation. The broadening in the DC case, is due to a decrease of lifetime in the spin blockade regime as described in Sec.~\ref{sec:DC-Hanle}. In the AC case, line broadening arises from increased scattering of electrons from the left electrode with the resonantly driven \ac{QD} state $|+\rangle\langle+|_\mathrm{rot}$ in an analogous fashion. In both cases, the line width is broadened by the DLT-driving strength captured by $\gamma_{n\mathrm{L}}(t)$.

\subsection{FLT vs. DLT dynamics}
So far, we have introduced phenomenological parameters capturing the FLT and DLT, both for the DC and the AC case summarized in Table~\ref{tab:param}. In this section, we focus on the differences between coherent and dissipative processes, namely on FLT- and DLT-driven EPR, providing explicit numerical results and examples highlighting the experimental signatures of spin torque dynamics in a realistic \ac{EPR} spectroscopy setup.
We will illustrate the role of the spin torques in driving distinct spectroscopic signatures and how the Hanle effect is closely related to the projective readout of the QD spin, as indicated by Table~\ref{tab:param}.

\begin{table}[ht]
\caption{Model parameters and their respective roles in driving and detection of the \ac{QD} spin. The detection mechanism of the spin state is particularly relevant for the emergence and shape of spectral resonance lines.
}
\begin{ruledtabular}\label{tab:param} 
\begin{tabular}{lll}
Parameter & Role in driving & Role in read-out\\
\hline
 $|\boldsymbol{B}_{\mathrm{xc},\mathrm{L}}^{(0)}|$ & DC-FLT & none \\
$|\boldsymbol{B}_{\mathrm{xc},\mathrm{L}}^{(1)}|$& AC-FLT &  none\\ 
 
 $\gamma_\mathrm{4L}^{(0)}$ & DC-DLT & DC Hanle effect \\ &&(DC read-out)\\
 
 $\gamma_\mathrm{4L}^{(1)}$ & DC- \& AC-DLT & DC \& AC Hanle effects \\ &&(DC \& homodyne read-out)
 
\end{tabular}
\end{ruledtabular}
\end{table}

\subsubsection{Spin torque tomography}

\begin{figure*}
    \centering
    \includegraphics[scale=1]{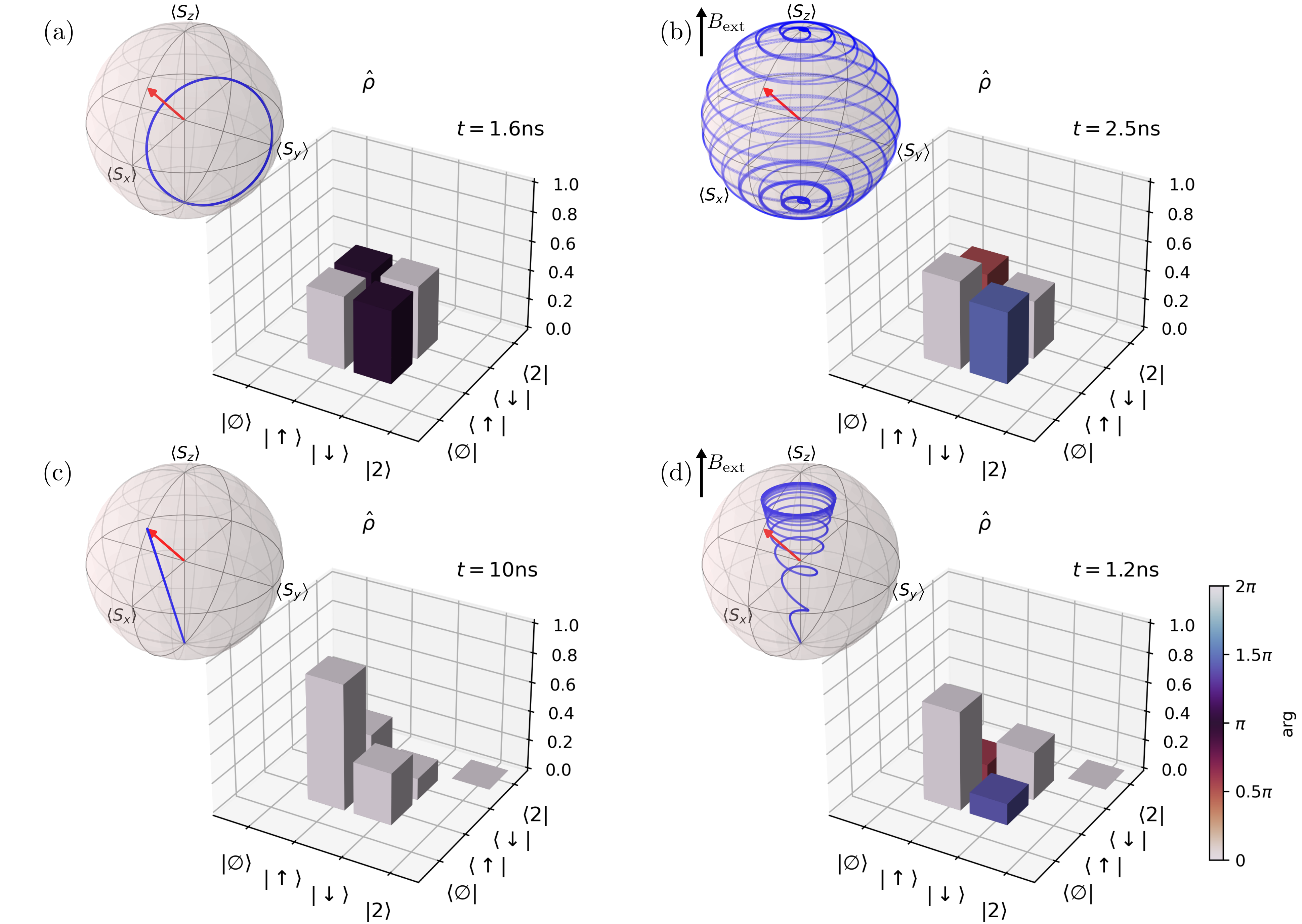}
   \caption{Comparison of spin and charge dynamics for DC- and AC-driven torques  (a) Larmor precession induced by the DC-FLT due to the electrode's exchange field. (b) Coherent \ac{EPR} induced by the AC-FLT in the absence of charge transfer (equivalent to $B_1$-driving). (c) Initialization of the dot spin along the spin polarization direction of the magnetic electrode induced by the DC-DLT in the presence of charge transfer. (d) \ac{EPR} driven by the AC-DLT synchronized to the Larmor frequency of the dot spin. The red arrows and blue lines in the insets represent the direction of the electrode's spin polarization and the trajectory of the dot spin, respectively. The plots represent the density matrix elements of the zero-, singly- and doubly-charged dot states at selected times $t$ (for (a) and (c), the density matrix approaches the maximal possible population inversion; in (b) and (d), states during a precession cycle are shown, when the transversal polarization of the QD points along $x$).}
    \label{fig:ACDC_B1B0}
\end{figure*}

Figure~\ref{fig:ACDC_B1B0} contrasts the FLT and DLT dynamics by showing the generalized Bloch trajectories of the dot spin in the different regimes. The case of static and oscillating exchange fields are shown in Figure~\ref{fig:ACDC_B1B0} (a)~and~(b), respectively, whereas the DC- and AC-DLT are shown in panels (c)~and~(d). In addition, Fig.~\ref{fig:ACDC_B1B0} illustrates snapshots of the quantum dot's density matrix. In the DLT case, charge fluctuations lead to occupation of the doubly charged dot state $|\hspace{0.15em}2\hspace{0.15em}\rangle$, while the exchange field dynamics involve only virtual charge states. 

As mentioned earlier, a crucial difference between the dynamics induced by the DLT and FLT is their coupling mechanism: the DLT involves spin-dependent charge transport, whereas the FLT acts exclusively on the spin degrees of freedom without affecting the charge states. This difference is illustrated in Figs.~\ref{fig:ACDC_B1B0}(b) and \ref{fig:ACDC_B1B0}(d). Fig.~\ref{fig:ACDC_B1B0} compares time-resolved snapshots of the reduced system density matrix $\hat{\rho}(t)$ initialized in the spin-down state $|\!\downarrow\,\rangle\langle\,\downarrow\!|$ in different driving scenarios. In the case of DLT driving, a small but finite occupation of the doubly charged state $|\hspace{0.15em}2\hspace{0.15em}\rangle\langle\hspace{0.15em}2\hspace{0.15em}|$ emerges, reflecting actual charge transfer.  In contrast, under AC-FLT driving, only the spin populations and coherences $\rho_{\sigma\sigma^\prime}$ with $\sigma,\sigma^\prime \in \{\uparrow,\downarrow\}$ are affected, with no occupation of charge states beyond the single-particle sector.

\subsubsection{High-harmonic driving}
Beyond the AC effects described above, periodic modulation of the junction bias can induce higher-order harmonics in the time dependence of the exchange field and charge-transfer rates. These arise because of intrinsic nonlinearities in the tunnel coupling and lead to deviations from simple sinusoidal dynamics. 
The Lindblad master equation Eq.~(\ref{eq:lindblad-me}) allows for direct comparison of the harmonic expansion of both the field-like contributions to $\hat{H}_\mathrm{xc}(t)$ and of the time-dependent rates $\gamma_{n\alpha}(t)$, analogous to the Floquet master equation treatment in Ref.~\onlinecite{reina-galvez_study_2021}. 
The general form of the resulting time-dependent quantities can be written as

\begin{align}\label{eq:higher}
    \boldsymbol{B}_{\mathrm{xc},\alpha}(t)=& \boldsymbol{B}_{\mathrm{xc},\alpha}^{(0)} + 
     \boldsymbol{B}_{\mathrm{xc},\alpha}^{(1)}\cos(\omega t) \nonumber\\
    & +\boldsymbol{B}_{\mathrm{xc},\alpha}^{(2)} \cos(2\omega t + \tilde{\varphi}^{(2)}_{\alpha}) + \dots\space,\\
    \gamma_{n\alpha} (t) =& \gamma_{n\alpha}^{(0)}+ 
    \gamma_{n\alpha}^{(1)}\frac{1}{2}\left( 1+\cos(\omega t+\varphi_{n\alpha}^{(1)})\right)
     \nonumber\\
    &+\gamma_{n\alpha}^{(2)}\frac{1}{2}\left( 1+\cos(2\omega t+\varphi_{n\alpha}^{(2)})\right) + \dots
    \,,
\end{align}
for phase shifts $\tilde{\varphi}_\alpha^{(2)},...,\varphi^{(1)}_{n\alpha},...$ with respect to 1st harmonic FLT driving. For the remainder of this paper, we assume no such phase shifts, corresponding to $\tilde{\varphi}_\alpha^{(k+1)}=\varphi^{(k)}_{n\alpha}=0$ (for all $k\geq1$).
The presence of these higher harmonics enables \ac{EPR} spectroscopy beyond the fundamental Larmor frequency \cite{schlitz_high_harmonic_2025}. For each harmonic $k$, a resonance condition $\omega_0 = k\omega$ can be satisfied, leading to spin locking via DLT at multiple frequencies. For DLT driving, the strength and linewidth of each resonance are governed by the corresponding harmonic amplitude $\gamma^{(k)}_{n\alpha}$, which determines the effective spin-transfer-induced relaxation time $T_\mathrm{D}$. 

\subsubsection{Dependence of the EPR signal on the orientation of the electrode's magnetization}
\begin{figure}
    \centering
    \includegraphics[scale=0.75]{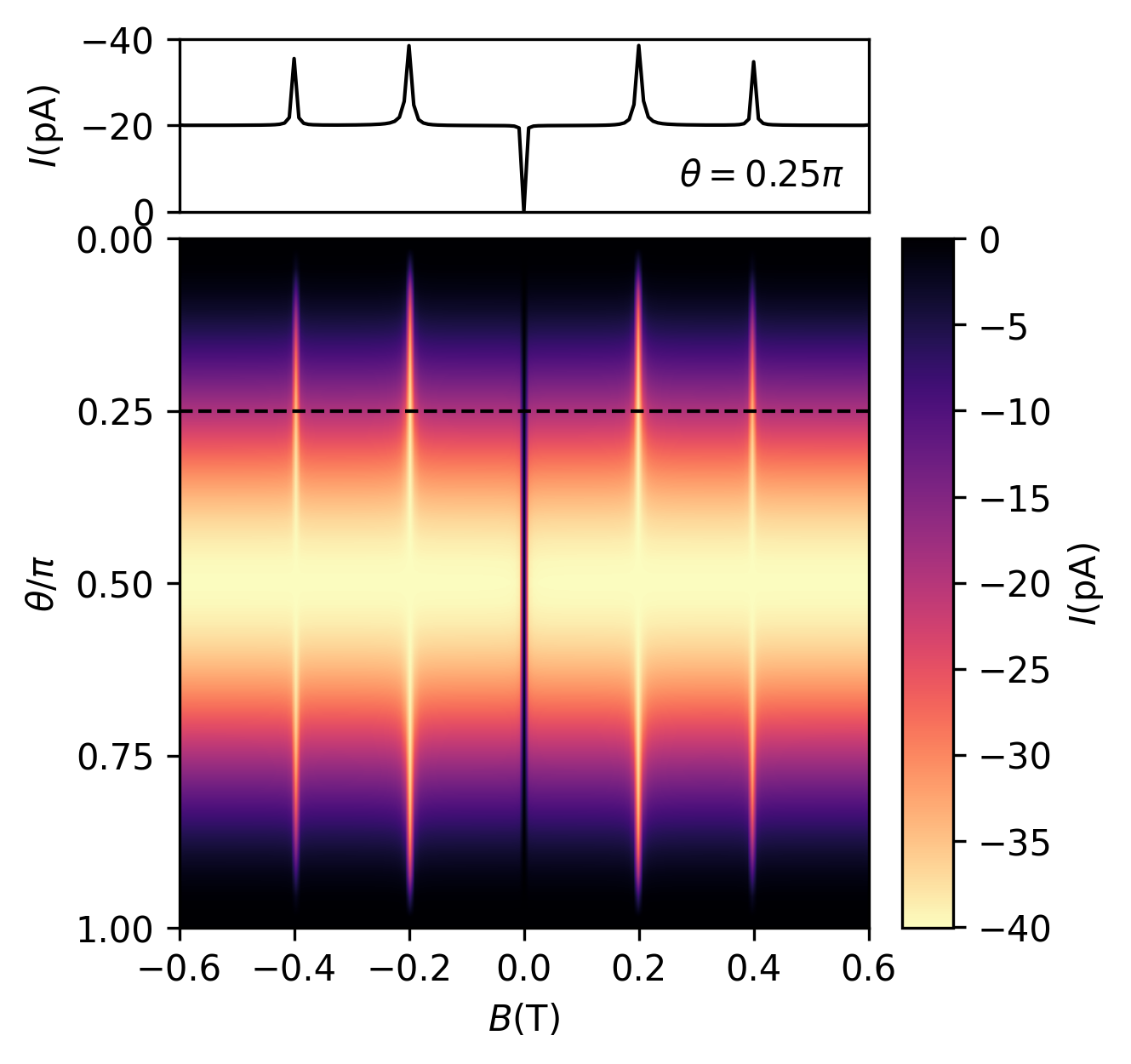}
    \caption{Tip angle dependence of dominantly AC-FLT-driven EPR signal. First and second harmonic driving is realized by $|\boldsymbol{B}_{\mathrm{xc},\mathrm{L}}^{(1)}|=0.0025\, \mathrm{rad}\!\cdot\!\mathrm{ps}^{-1}\cdot\hbar/(g_\mathrm{s}\mu_\mathrm{B})\approx 0.014\;\mathrm{T}$ and $|\boldsymbol{B}_{\mathrm{xc},\mathrm{L}}^{(2)}|=0.00125\,\mathrm{rad}\!\cdot\!\mathrm{ps}^{-1}\cdot\hbar/(g_\mathrm{s}\mu_\mathrm{B})$, respectively.
    The spin-state is read out with a small DC corresponding to sequential tunneling with $\gamma_{4\mathrm{L}}^{(0)}= 0.0005\, \mathrm{rad}\!\cdot\!\mathrm{ps}^{-1}$ and $\gamma_{6\mathrm{R}}^{(0)}, \gamma_{8\mathrm{R}}^{(0)} = 0.5\, \mathrm{rad}\!\cdot\!\mathrm{ps}^{-1} \approx 0.33\;\mathrm{meV}/\hbar$.}
    \label{fig:map2D_angle_B1}
\end{figure}

To further highlight the distinct signatures of AC-FLT-driven and DLT-driven \ac{EPR}, we present resonance spectra up to second-harmonic as a function of the polarization angle $\theta$ of the left electrode, i.e., of the magnetic tip in an STM junction, relative to the $z$-axis.
Figure~\ref{fig:map2D_angle_B1} shows the case of AC-FLT-driving, described by 
\begin{equation}\label{eq:Hxc_2omega}
   \boldsymbol{B}_{\mathrm{xc},\alpha}(t) = \boldsymbol{B}_{\mathrm{xc},\alpha}^{(1)} \cos(\omega t)+ \boldsymbol{B}_{\mathrm{xc},\alpha}^{(2)} \cos(2\omega t)\,,
\end{equation}
with a small DC-DLT enabling the spin-selective transport necessary to read out the spin-state.
Here, the driving is purely field-like, and the resonance signal requires sequential transport to be visible. This is enabled by finite hopping rates to and from the dot, $\gamma^{(0)}_{n,\mathrm{L}}$ and $\gamma^{(0)}_{n,\mathrm{R}}$. Detection via the \ac{QD} current under DC conditions results in symmetric line shapes, consistent with the findings of Ref.~\onlinecite{kovarik_spin_2024}. Notably, the zero-field Hanle resonance and the first- and second-harmonic lines exhibit opposite signs, reflecting their distinct dynamical origins. The components $\boldsymbol{B}_{\mathrm{xc},\alpha}^{(1)}$ and $\boldsymbol{B}_{\mathrm{xc},\alpha}^{(2)}$ counteract dynamic spin locking, which would otherwise suppress the conductance at resonance. 

\begin{figure}
    \centering
    \includegraphics[scale=0.75]{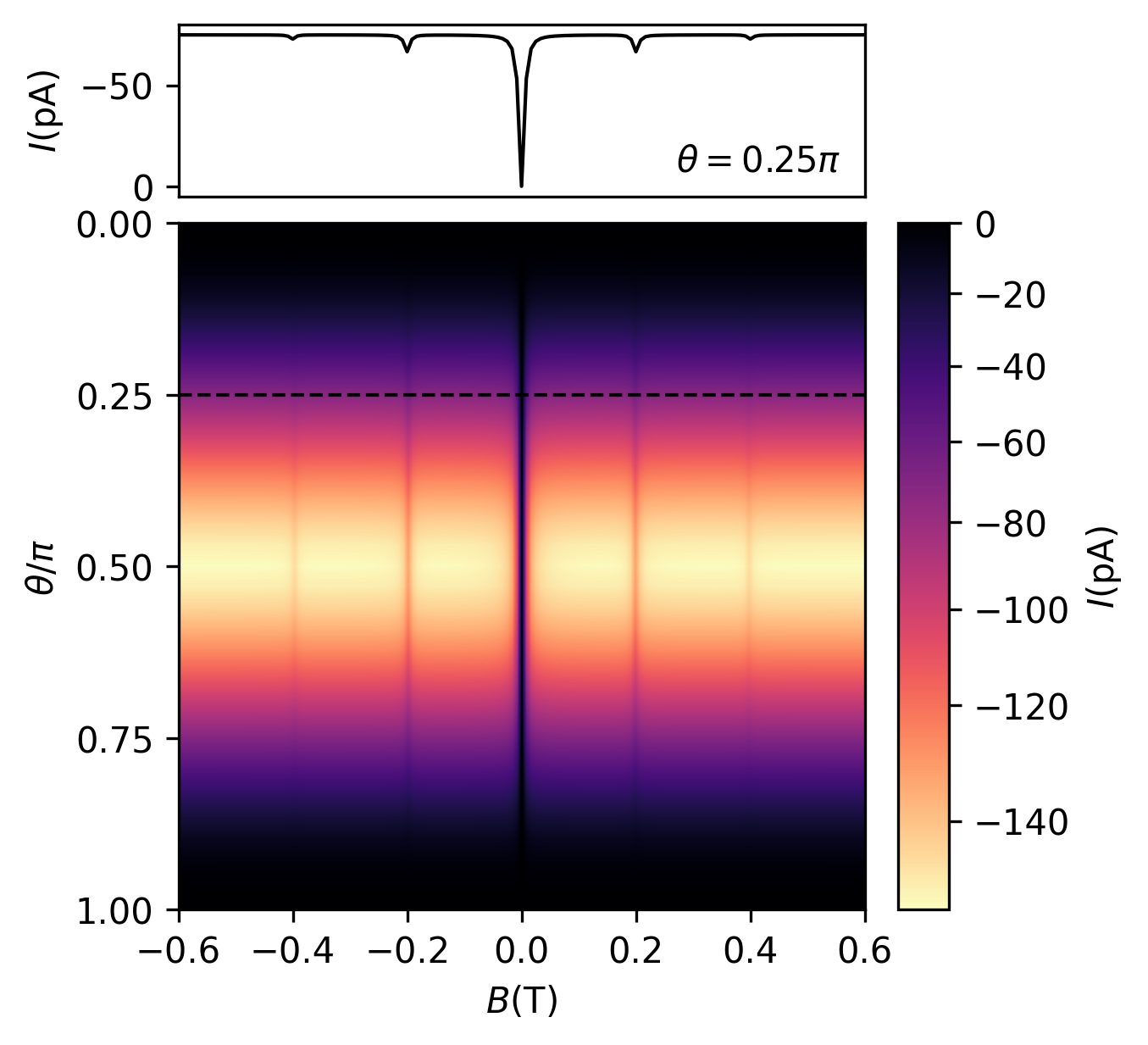}
    \caption{Polarization angle $\theta$ dependence of DLT-driven EPR signal in the case of first and second harmonic driving. The relevant transport parameters are
    $\gamma_{4\mathrm{L}}^{(1)}= 0.0025\, \mathrm{rad}\!\cdot\!\mathrm{ps}^{-1}$, $\gamma_{4\mathrm{L}}^{(2)}= 0.00125\, \mathrm{rad}\!\cdot\!\mathrm{ps}^{-1}$ and $\gamma_{6\mathrm{R}}^{(0)}, \gamma_{8\mathrm{R}}^{(0)} = 0.5\, \mathrm{rad}\!\cdot\!\mathrm{ps}^{-1} \approx 0.33\;\mathrm{meV}/\hbar$.
    }
    \label{fig:map2D_angle_STT}
\end{figure}

In contrast, for pure DLT driving, the resonance lines appear as conductance dips, similar to the Hanle resonance at zero field as shown in Fig.~\ref{fig:map2D_angle_STT}.
The similarities between DC- ("0th harmonic"), and AC- (1st and 2nd harmonics) DLT driving, highlight the common underlying principle. The following section further addresses the detection via the Hanle effect.

\subsubsection{Hanle effects due to the DC- and AC-DLT}
We now focus on the different harmonics of pure DLT-driving. As discussed in Sec.~\ref{sec:AC-STT-QD}, the Hanle feature at zero field is due to the DC-DLT or, equivalently, to the 0th harmonic component of an AC-DLT. 
The DLT-induced EPR peaks where $\omega_\mathrm{0}$ is a multiple of the driving frequency $\omega$~\cite{schlitz_high_harmonic_2025} correspond to a high-harmonic generalization of the Hanle effect.

\begin{figure}
    \centering
    \includegraphics[scale=0.75]{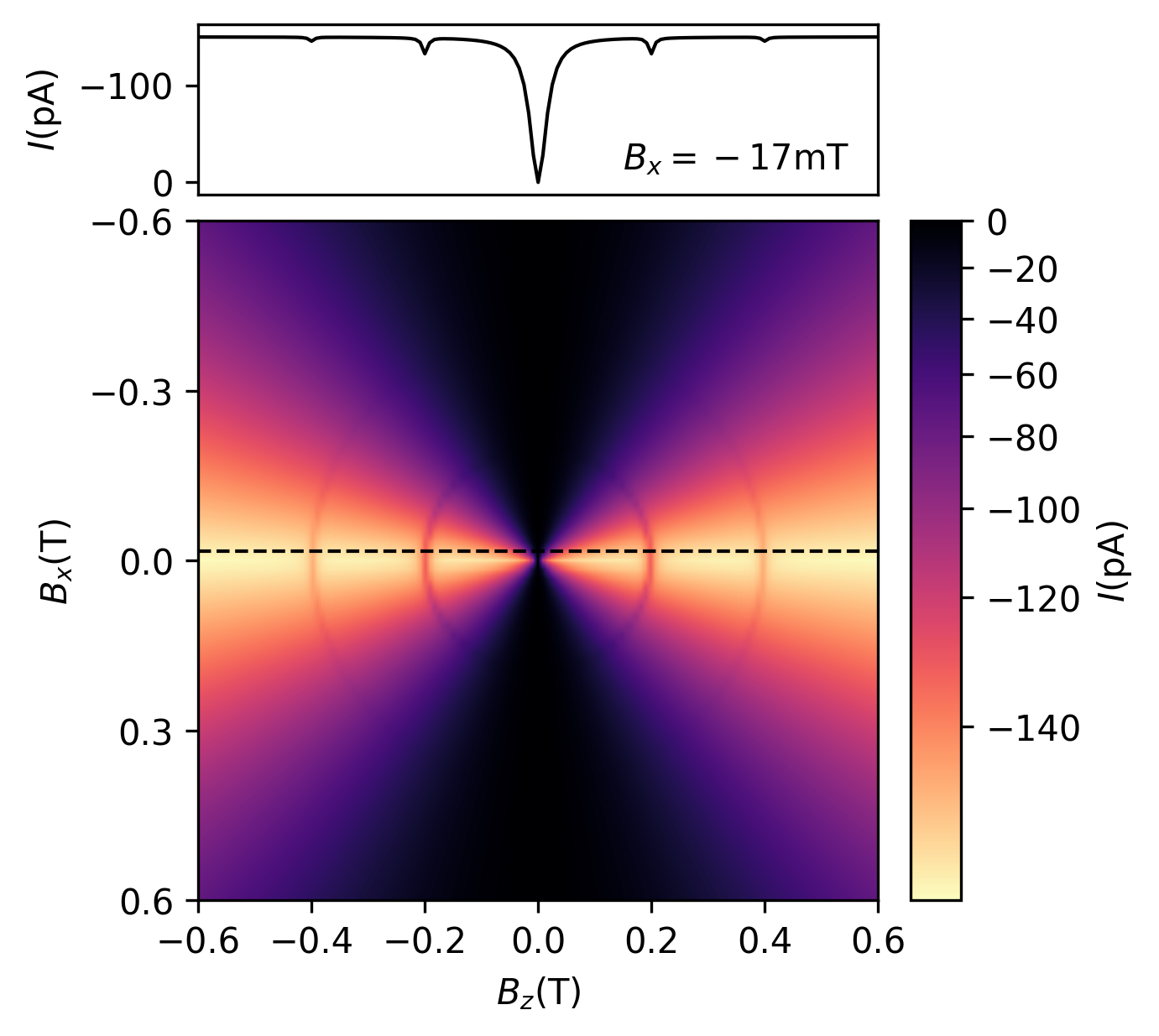}
    \caption{Conductance peak at zero magnetic field (DC Hanle effect) and first- and second-harmonic DLT-driven EPR lines (AC Hanle effect) at 100~mT and 200~mT away from the origin $B_x, B_z = 0$.
    The transport rates relevant for driving and read-out are
    $\gamma_{4\mathrm{L}}^{(1)}= 0.0025\, \mathrm{rad}\!\cdot\!\mathrm{ps}^{-1}$, $\gamma_{4\mathrm{L}}^{(2)}= 0.00125\, \mathrm{rad}\!\cdot\!\mathrm{ps}^{-1}$ and $\gamma_{6\mathrm{R}}^{(0)}, \gamma_{8\mathrm{R}}^{(0)} = 0.5\, \mathrm{rad}\!\cdot\!\mathrm{ps}^{-1} \approx 0.33\;\mathrm{meV}/\hbar$.
    }
    \label{fig:mapBx}
\end{figure}

In Fig.~\ref{fig:mapBx}, we show a vector magnetic field map of the DC and AC Hanle effects up to second harmonic driving. A key distinction between the DC Hanle effect, also reported in Ref.~\onlinecite{busz_hanle_2023}, and the AC Hanle effects lies in their response to a field component $B_x$ aligned with the electrode's polarization. In the DC case, the line broadens linearly with $B_x$, which is not observed in the AC case. Furthermore, under AC conditions, resonance occurs when the total effective field satisfies $\sqrt{B_x^2+B_z^2} \propto \omega_0$, resulting in ring-shaped resonance features that are concentric around the origin. These rings mark the resonance condition for spin locking under harmonic driving. We identify this build up of a partial blockade at magnetic resonance conditions as the AC Hanle effect. 
 Although DC- and AC-DLT have fundamentally analog descriptions, their spectroscopic character behaves differently under a transverse magnetic field. This leads to scenarios in which the line-width of the DC Hanle feature drastically differs from the EPR line width in the AC case. Furthermore, the apparent shifting of the EPR lines due to $B_x$ needs to be taken into account when interpreting spectral data in which typically an external field $B_z$ is swept \cite{seifert_single-atom_2020}. Moreover, the dependence of the spectral  line position on $B_x$ differs between the 1st and 2nd harmonic EPR lines.

\subsubsection{DC and homodyne detection in driven sensor-spectator system.}
As a final example, we consider a coupled sensor-spectator \ac{QD} system subject to both AC-FLT and AC-DLT-driven \ac{EPR}. This configuration combines field-driven and spin-torque-driven dynamics to reveal interference effects between different transport pathways.

\begin{figure}
    \centering
    \includegraphics[scale=0.75]{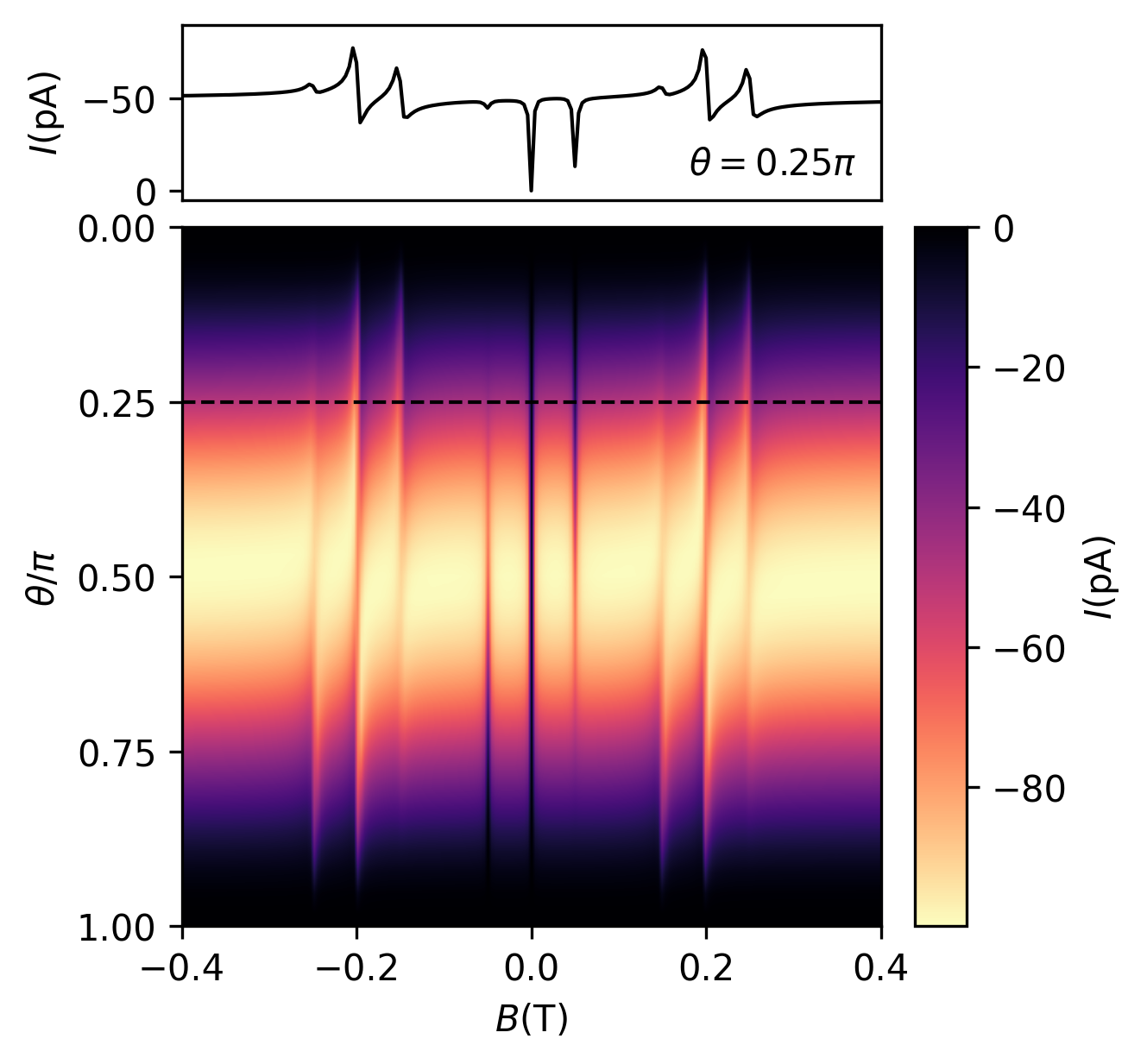}
    \caption{ZFF and EPR sidebands in the sensor-spectator system due to exchange coupling under AC-FLT and DLT driving. The coupling is given by $J=0.05\,g_\mathrm{s} \mu_\mathrm{B}/2\approx0.0029 \;\mathrm{meV}$
    The left dot is simultaneously driven by an FLT and a DLT with parameters  $|\boldsymbol{B}_{\mathrm{xc},\mathrm{L}}^{(1)}|=0.0025\, \mathrm{rad}\!\cdot\!\mathrm{ps}^{-1}\cdot\hbar/(g_\mathrm{s}\mu_\mathrm{B})$, $\gamma_{4\mathrm{L}}^{(1)}= 0.0025\, \mathrm{rad}\!\cdot\!\mathrm{ps}^{-1}$ and $\gamma_{6\mathrm{R}}^{(0)}, \gamma_{8\mathrm{R}}^{(0)} = 0.5\hbar\, \mathrm{rad}\!\cdot\!\mathrm{ps}^{-1} \approx 0.33\;\mathrm{meV}$.
    Asymmetry of the resonance profiles indicates the presence of AC-detection (homodyning) of the precessing spin slightly detuned from the resonance position or the first harmonic side-bands close to $\pm0.2\;\mathrm{T}$.
    }
    \label{fig:mapcoupled}
\end{figure}

Figure~\ref{fig:mapcoupled} shows the resulting current map, where the central zero-field feature of the DC Hanle effect is accompanied by two spectral lines at magnetic fields of $\pm 0.2$~T, corresponding to the EPR resonance conditions. Furthermore, for all three lines, sidebands arise from exchange interactions between the sensor and spectator spins and reflect both DC- \cite{mcmillan_image_2020} and AC-driven dynamics \cite{reina-galvez_study_2021}. 

The line-profile of six EPR lines (including side bands) are no longer described by a symmetric Lorentzian shape as was the case in Fig.~\ref{fig:map2D_angle_STT}, for example. The spectral profile shown in top of Fig.~\ref{fig:mapcoupled} exhibit both symmetric and asymmetric lineshape contributions, which arise from combined AC-FLT and DLT driving \cite{kovarik_spin_2024}.

For pure DLT-driven coherent precession, read-out of the oscillatory terms of the form $\cos(\omega_0t){\hat{S}_x}/{\hbar}$ (see Eq.~(\ref{eq:precessing_states}) ) is achieved by homodyne detection. The underlying principle is a down-conversion of transversal spin components oscillating with $\omega_0$, by frequency matching the applied AC electro-magnetic field such that $\omega=\omega_0$. The result is a mixing of the precessing spin-components with $\gamma^{(1)}_{n\alpha}\frac{1}{2}\cos(\omega t)$ in Eq.~(\ref{eq:gamma}) leading to a detectable DC conductance in the QD current \cite{kovarik_spin_2024}. This leads to time-averaged terms $\propto\overline{ \cos(\omega_0t)\!\cdot\!\cos(\omega t)}$, which only contribute to the DC current signal when the absolute detuning $|\omega-\omega_0|$ is within the spectroscopic linewidth, proportional to $\gamma^{(1)}_{n\mathrm{L}}$.

If the AC-drive is dominantly field-like, in contrast, homodyne detection leads to an asymmetric line shape \cite{ast_theory_2024,bae_enhanced_2018}. This can be explained by an AC-FLT-driving and an AC-DLT-readout scenario, in which the steady-state precession is dominated by an oscillatory component $\propto\cos(\omega_0t+\pi/2){\hat{S}_x}$, which exactly cancels at $\omega=\omega_0$ when mixing with $\gamma^{(1)}_{n\alpha}\frac{1}{2}\cos(\omega t)$ and requires a finite detuning $\omega\neq\omega_0$ to yield a spectroscopic signal. 
Since both driving and read-out depend on the polarization vector of the left electrode, the spectroscopic signal depends on the polarization angle $\theta$. Towards $\theta=\pi/2$ the asymmetric contribution due to AC-FLT-driving becomes indistinguishable from the background, as expected for an asymmetric profile that switches sign and thus crosses zero. The zero-field line at $B=0$ and its two side bands, in contrast, do not exhibit the dominant asymmetries of their EPR counterparts and are in agreement with the DC description in Sec.~\ref{sec:dimer-collapse} in the limit of spectral separability of the DC and AC line profiles.

\begin{figure}
    \centering
    \includegraphics[scale=0.75]{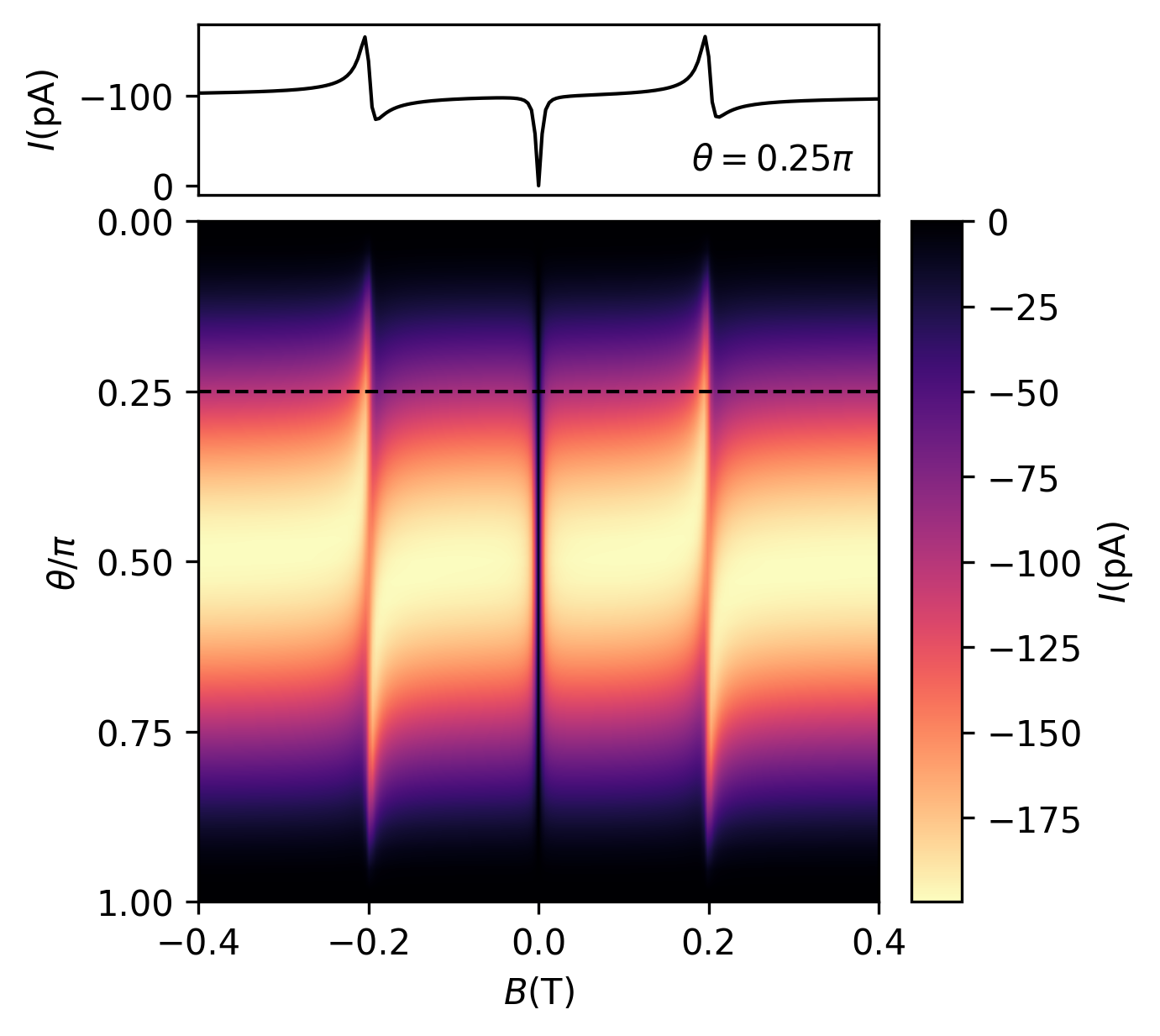}
    \caption{Collapse of zero-field Hanle feature and EPR sidebands due to stochastic tunneling through both \acp{QD} simultaniously in the sensor-spectator system.
    Both dots are driven by the same FLT as in Fig.~\ref{fig:mapcoupled} and DLT with parameters  $|\boldsymbol{B}_{\mathrm{xc},\mathrm{L}}^{(1)}|=0.0025\, \mathrm{rad}\!\cdot\!\mathrm{ps}^{-1}\cdot\hbar/(g_\mathrm{s}\mu_\mathrm{B})$, $\gamma_{4\mathrm{L}}^{(1)}= 0.0025\, \mathrm{rad}\!\cdot\!\mathrm{ps}^{-1}$ and $\gamma_{6\mathrm{R}}^{(0)}, \gamma_{8\mathrm{R}}^{(0)} = 0.5\, \mathrm{rad}\!\cdot\!\mathrm{ps}^{-1} \approx 0.33\;\mathrm{meV}/\hbar$. The current is computed only through the left \ac{QD} for simplicitly. The current through both dots is given by twice the single channel current due to symmetry.
    }
    \label{fig:mapcollapse}
\end{figure}

In Fig.~\ref{fig:mapcollapse}, we show how these EPR sidebands vanish when transport occurs by simultaneous tunneling through both \acp{QD}. This behavior, discussed in Sec.~\ref{sec:dimer-collapse}, represents a form of quantum interference or “collapse” of distinct signal paths, where the coherent sideband structure is lost due to stochastic tunneling correlations. The total current is computed in analogy to Sec.~\ref{sec:dimer-collapse}. Notably, this suppression of resonance features occurs in both the DC- and AC-driven regimes, underscoring a fundamental equivalence in their transport signatures under such coupled conditions.

\section{Conclusions}
We investigated spin-selective charge transport in \ac{QD} systems using a Lindblad-type master equation formalism, building upon the framework established in Ref.~\onlinecite{braun_theory_2004} and \onlinecite{reina-galvez_study_2021}. This approach captures the interplay between unitary spin dynamics and stochastic charge transfer processes, and enables a complete real-time description of both charge populations and spin coherences.

In the absence of an external magnetic field, we recover the emergence of a zero-field conductance peak, the so-called DC Hanle effect, arising from spin initialization via spin-selective tunneling, as originally proposed in Ref.~\onlinecite{braun_hanle_2005}. We show that this feature persists even when introducing a time-dependent modulation of the tunneling rates, representing the essential mechanism of spin-torque-driven electron paramagnetic resonance (DLT-\ac{EPR}). At resonance conditions, when the modulation frequency matches the Larmor frequency, we observe distinct conductance dips due to dynamic spin locking.

By extending the model to include time-dependent Hamiltonian terms, such as oscillating exchange fields (AC-FLT-driving), we highlight the contrasting signatures between unitary and dissipative driving mechanisms. In particular, while DLT-driven \ac{EPR} leads to symmetric peaks in the conductance spectrum, field-driven (AC-FLT) resonance produces Lorentzian lines of opposite sign due to homodyne detection and the absence of spin-dependent charge transfer. This distinction is further emphasized in angle-dependent spectra, where DC- and AC-FLT-driven responses exhibit qualitatively different dependencies of the line shape on $\theta$ between the polarization vector of the left electrode and $\boldsymbol{B}_\mathrm{ext}$.
 
We also explored the role of higher harmonics in both the tunnel coupling modulation and the exchange field, revealing that second-harmonic driving leads to higher-order \ac{EPR} sidebands. Each harmonic corresponds to a distinct spin-locking condition $\omega_0 = k\omega$, and the associated spin relaxation rate is determined by the strength of the corresponding harmonic in $\gamma^{(k)}_{n\alpha}$.

Finally, we examined the coupled sensor-spectator dot system and demonstrated how DC and AC torque-driving give rise to sidebands. These features are sensitive to coherent exchange interactions and may disappear when transport occurs simultaneously through both \acp{QD}. Such transport-induced collapse of resonance features highlights the importance of pathway control in coupled spintronic systems.

Our model is useful for interpreting recent experimental developments in spin-dependent transport spectroscopy, particularly in scanning tunneling microscopy of magnetic atoms, molecular spin systems, and gate-tunable \acp{QD}. It also provides a theoretical foundation for time-resolved spectroscopy protocols in hybrid nanostructures, including carbon nanotubes, 2D materials, and molecular break junctions. Finally, our results pave the way toward exploring non-linear spin dynamics, time-domain control of spin torque, and quantum metrology schemes based on harmonic driving and coherent spin transfer.

\begin{acknowledgments}
We wish to thank J. Reina-G\'alvez, M. Flatt\'e, N. Lorente, and S. R. McMillan for helpful discussions.

This work was supported by the Swiss National Science Foundation (Grant No. CRSII5\_205987) and by ETH (Grant No. ETH-45 20-2).
\end{acknowledgments}

\section*{Data availability}
The code to generate the data in the figures of this article is openly available \cite{paperdata}.

\vspace{2em} 

\input{appendices}

\FloatBarrier
\bibliography{bib}

\end{document}

%% file: diagrams/scatter1.tex
  
  
        
    
    
    




\begin{figure}[ht]
\centering
\vspace{1em}
\begin{minipage}{1.0em}
\vspace{-8em}(a)
\end{minipage}
\begin{minipage}{11em}
\begin{fmffile}{diagram4}  
\begin{fmfgraph*}(90,40)
    \fmfleft{i1,i2} 
    \fmfright{o1,o2}
    \fmf{plain_arrow,tension=5/6}{i2,v2}
    \fmf{plain_arrow,tension=1/6}{v2,o2}
    \fmf{plain_arrow,tension=5/6}{o1,v1}
    \fmf{plain_arrow,tension=1/6}{v1,i1}
    \fmffreeze
    \fmf{dashes_arrow,label=L,label.side=right}{v1,v2}    
    \fmflabel{$\uparrow$}{i2}
    \fmflabel{$\uparrow$}{i1}
    \fmflabel{$2$}{o2}
    \fmflabel{$2$}{o1}
  \end{fmfgraph*}
\end{fmffile}
\end{minipage}
\hfill
\begin{minipage}{1.0em}
\vspace{-8em}(b)
\end{minipage}
\begin{minipage}{11em}
\begin{fmffile}{diagram5}  
\begin{fmfgraph*}(90,40)
    \fmfleft{i1,i2} 
    \fmfright{o1,o2}
    \fmf{plain_arrow,tension=5/6}{i2,v2}
    \fmf{plain_arrow,tension=1/6}{v2,o2}
    \fmf{plain_arrow,tension=5/6}{o1,v1}
    \fmf{plain_arrow,tension=1/6}{v1,i1}
    \fmffreeze
    \fmf{dashes_arrow,label=R,label.side=right}{v2,v1}    
    \fmflabel{$2$}{i2}
    \fmflabel{$2$}{i1}
    \fmflabel{$\downarrow$}{o2}
    \fmflabel{$\downarrow$}{o1}
  \end{fmfgraph*}
\end{fmffile}
\end{minipage}

\vspace{6em}

\begin{minipage}{1.0em}
\vspace{-7em}(c)
\end{minipage}
\begin{minipage}{11em}
\begin{fmffile}{diagram6}  
    \begin{fmfgraph*}(90,40)
    \fmfleft{i1,i2} 
    \fmfright{o1,o2}
    \fmf{plain_arrow,tension=5/6}{i2,v1}
    \fmf{plain_arrow,tension=1/4,label=$2$,label.side=left}{v1,v2}
    \fmf{plain_arrow,tension=5/6}{v2,o2}
    \fmf{plain_arrow}{o1,i1}
    \fmffreeze
    \fmf{dashes_arrow,label=L,label.side=left,left=0.5}{v2,v1}
    \fmflabel{$\uparrow$}{i2}
    \fmflabel{$\uparrow$}{i1}
    \fmflabel{$\downarrow$}{o2}
    \fmflabel{$\uparrow$}{o1}
  \end{fmfgraph*}
\end{fmffile}
\end{minipage}
\hfill
\begin{minipage}{1.0em}
\vspace{-7em}(d)
\end{minipage}
\begin{minipage}{11em}
  \begin{fmffile}{diagram7}
  \begin{fmfgraph*}(90,40)
    \fmfleft{i1,i2} 
    \fmfright{o1,o2}
    \fmf{plain_arrow,tension=5/6}{v1,i1}
    \fmf{plain_arrow,tension=1/4,label=$2$,label.side=left}{v2,v1}
    \fmf{plain_arrow,tension=5/6}{o1,v2}
    \fmf{plain_arrow}{i2,o2}
    \fmffreeze
    \fmf{dashes_arrow,label=L,label.side=left,left=0.5}{v1,v2}
    \fmflabel{$\uparrow$}{i2}
    \fmflabel{$\uparrow$}{i1}
    \fmflabel{$\uparrow$}{o2}
    \fmflabel{$\downarrow$}{o1}
  \end{fmfgraph*}
\end{fmffile}
\end{minipage}

    \vspace{2em}
    \caption{Lowest order scattering diagrams describing the quantum dot evolution due to tunnel coupling to electrodes $\mathrm{L}$~or~$\mathrm{R}$. (a) Charging process $|\!\uparrow\,\rangle\langle\,\uparrow\!| \overset{\Gamma_\mathrm{L}}{\longrightarrow}|\hspace{0.15em}2\hspace{0.15em}\rangle\langle\hspace{0.15em}2\hspace{0.15em}|$ for a singly charged dot initially in the $|\!\uparrow\,\rangle\langle\,\uparrow\!|$ configuration, which acquires an electron with spin-$\downarrow$  from the left electrode and leaves the dot in the doubly charged state $|\hspace{0.15em}2\hspace{0.15em}\rangle\langle\hspace{0.15em}2\hspace{0.15em}|$. (b) An initially doubly charged dot loses a spin-$\uparrow$ electron to the right electrode which corresponds to a process 
$|\hspace{0.15em}2\hspace{0.15em}\rangle\langle\hspace{0.15em}2\hspace{0.15em} | \overset{\Gamma_\mathrm{R}}{\longrightarrow}|\!\downarrow\,\rangle\langle\,\downarrow\! |$. (c) Virtual charging from the left electrode corresponding to $|\!\uparrow\,\rangle\langle\,\uparrow\! | \overset{\Gamma_\mathrm{L}}{\longrightarrow}|\!\downarrow\,\rangle\langle\,\uparrow\! |$, and (d) $|\!\uparrow\,\rangle\langle\,\uparrow\! | \overset{\Gamma_\mathrm{L}}{\longrightarrow}|\!\uparrow\,\rangle\langle\,\downarrow\! |$.}

        \label{fig:scatter4}
\end{figure}




%% file: diagrams/scatter-virt.tex
\newcommand{\eqgraph}[3]{%
  \begin{gathered}
  \raisebox{0pt}[\dimexpr\height+#1][\dimexpr\depth+#2]{\ignorespaces#3\unskip}%
  \end{gathered}
}

\begin{fmffile}{diagram2}
  \begin{equation}
    \Sigma_{\downarrow\uparrow ,\uparrow \uparrow} = 
    \mspace{20mu}
    \eqgraph{1ex}{1ex}{
    \begin{fmfgraph*}(60,35)
    \fmfleft{i1,i2} 
    \fmfright{o1,o2}
        
    \fmf{plain_arrow,label=$\emptyset$,label.side=left}{i2,o2}
    \fmf{plain_arrow}{o1,i1}
    
    \fmffreeze
    \fmf{dashes_arrow,right=0.5,label=L,label.side=right}{i2,o2}
    
    \fmflabel{$\uparrow$}{i2}
    \fmflabel{$\uparrow$}{i1}
    
    \fmflabel{$\downarrow$}{o2}
    \fmflabel{$\uparrow$}{o1}
    
  \end{fmfgraph*}
    }
    \quad
    +
    \quad
    \eqgraph{1ex}{1ex}{
    \begin{fmfgraph*}(60,35)
    \fmfleft{i1,i2} 
    \fmfright{o1,o2}
        
    \fmf{plain_arrow,label=2,label.side=left}{i2,o2}
    \fmf{plain_arrow}{o1,i1}
    
    \fmffreeze
    \fmf{dashes_arrow,left=0.5,label=L,label.side=left}{o2,i2}
    
    \fmflabel{$\uparrow$}{i2}
    \fmflabel{$\uparrow$}{i1}
    
    \fmflabel{$\downarrow$}{o2}
    \fmflabel{$\uparrow$}{o1}
    
  \end{fmfgraph*}
    }
  \quad, \label{eq:8}\end{equation}
  
\end{fmffile}

%% file: appendices.tex
\appendix

\section{Charge and spin quantum dot equations of motion}\label{app:eom}
Here, we shortly review the quantum dot equation of motion demonstrated in Ref.~\cite{braun_theory_2004} and how these relate to the collapse operators introduced in Eqs.~(\ref{eq:cops12})--(\ref{eq:cops78}) in Sec.~
\ref{sec:Lindblad-me}. The dynamics are fully captured by three charge and three spin equations of motions given by 
\begin{widetext}
\begin{align}
\frac{\text{d}}{\text{d}t}
\begin{pmatrix}
\rho_{\emptyset\emptyset} \\
\rho_{\uparrow\uparrow}+\rho_{\downarrow\downarrow} \\
\rho_{22}
\end{pmatrix}
=&
\sum_{\alpha = \mathrm{L},\mathrm{R}} \frac{\Gamma_\alpha}{\hbar}
\begin{pmatrix}
-2f_\alpha^\mathrm{e}(\epsilon) & f_\alpha^\mathrm{h}(\epsilon) & 0 \\
2f_\alpha^\mathrm{e}(\epsilon) & -f_\alpha^\mathrm{h}(\epsilon) - f_\alpha^\mathrm{e}(\epsilon + U) & 2f_\alpha^\mathrm{h}(\epsilon + U) \\
0 & f_\alpha^\mathrm{e}(\epsilon + U) & -2f_\alpha^\mathrm{h}(\epsilon + U)
\end{pmatrix}
\begin{pmatrix}
\rho_{\emptyset\emptyset} \\
\rho_{\uparrow\uparrow}+\rho_{\downarrow\downarrow} \\
\rho_{22}
\end{pmatrix}\notag\\
&+
\sum_{r = \text{L,R}} \frac{2p_\alpha\Gamma_\alpha}{\hbar}
\begin{pmatrix}
f_\alpha^\mathrm{h}(\epsilon) \notag\\
-f_\alpha^\mathrm{h}(\epsilon) + f_\alpha^\mathrm{e}(\epsilon + U) \notag\\
-f_\alpha^\mathrm{e}(\epsilon + U)
\end{pmatrix}
\boldsymbol{S} \cdot \boldsymbol{n}_\alpha \,,\\
\label{eq:occuptations}
\end{align}

and 

\begin{equation}
\frac{\text{d}\boldsymbol{S}}{\text{d}t} = \left(\frac{\text{d}\boldsymbol{S}}{\mathrm{d}t}\right)_{\mathrm{D}} + \left(\frac{\text{d}\boldsymbol{S}}{\mathrm{d}t}\right)_{\mathrm{S}} + \left(\frac{\text{d}\boldsymbol{S}}{\mathrm{d}t}\right)_{\mathrm{F}}\,,
\label{eq:spins}
\end{equation}
where
\begin{align*}
\left( \frac{\text{d}\boldsymbol{S}}{\text{d}t} \right)_{\mathrm{D}} &= \sum_{\alpha = \mathrm{L},\mathrm{R}} \frac{p_\alpha \Gamma_\alpha}{\hbar} 
\left[
f_\alpha^\mathrm{e}(\epsilon) \rho_{\emptyset\emptyset} + \frac{-f_\alpha^\mathrm{h}(\epsilon) + f_\alpha^\mathrm{e}(\epsilon + U)}{2} \left(\rho_{\uparrow \uparrow}+\rho_{\downarrow \downarrow}\right) - f_\alpha^\mathrm{h}(\epsilon + U) \rho_{22}
\right] \boldsymbol{n}_\alpha\,,\\
\left( \frac{\text{d}\boldsymbol{S}}{\text{d}t} \right)_{\mathrm{S}} &= -\sum_{\alpha = \mathrm{L},\mathrm{R}} \frac{\Gamma_\alpha}{\hbar} 
\left[
f_\alpha^\mathrm{h}(\epsilon) + f_\alpha^\mathrm{e}(\epsilon + U)
\right] \boldsymbol{S}\,,\\
\left( \frac{\text{d}\boldsymbol{S}}{\text{d}t} \right)_{\mathrm{F}} &= \boldsymbol{S} \times \sum_{\alpha = \mathrm{L},\mathrm{R}} \frac{g_\mathrm{s}\mu_\mathrm{B}}{\hbar}\boldsymbol{B}_{\mathrm{xc,\alpha}}\,.
\end{align*}
The evolution of the three spin components $\frac{\text{d}\boldsymbol{S}}{\text{d}t}$ are expressed by a damping-like torque contribution $\left( \frac{\text{d}\boldsymbol{S}}{\text{d}t} \right)_{\mathrm{D}}$ due to accumulation of spin along the unit vector $\boldsymbol{n}_\mathrm{\alpha}$ for electrode $\alpha$, a spin-relaxation term $\left( \frac{\text{d}\boldsymbol{S}}{\text{d}t} \right)_{\mathrm{S}}$ which allows the norm $|\boldsymbol{S}|$ to decay exponentially, and the field-like rotation $\left( \frac{\text{d}\boldsymbol{S}}{\text{d}t} \right)_{\mathrm{F}}$ with respective exchange fields $\boldsymbol{B}_{\mathrm{xc,\alpha}}$.

The equivalence between the Lindblad master equation and Eqs.~(\ref{eq:occuptations})~and~(\ref{eq:spins}) are found by computing the expressions $\cfrac{\mathrm{d}}{\mathrm{d}t}\mathrm{tr}\left\lbrace\vphantom{\frac{}{}} \,|\hspace{0.15em}\emptyset\hspace{0.15em}\rangle\langle\hspace{0.15em}\emptyset\hspace{0.15em}|\,\hat{\rho}(t)\right\rbrace$, $\cfrac{\mathrm{d}}{\mathrm{d}t}\mathrm{tr}\left\lbrace\left(\vphantom{\frac{}{}}|\!\uparrow\,\rangle\langle\,\uparrow\!|+|\!\downarrow\,\rangle\langle\,\downarrow\!|\,\right)\hat{\rho}(t)\right\rbrace$ and $\cfrac{\mathrm{d}}{\mathrm{d}t}\mathrm{tr}\left\lbrace\vphantom{\frac{}{}}\,|\hspace{0.15em}2\hspace{0.15em}\rangle\langle\hspace{0.15em}2\hspace{0.15em}|\,\hat{\rho}(t)\right\rbrace$ for the change in electron occupations, and $\cfrac{\mathrm{d}}{\mathrm{d}t}\mathrm{tr}\left\lbrace\vphantom{\frac{}{}} \,\hat{\sigma}_i\,\hat{\rho}(t)\right\rbrace \hbar/2$ for $i=x,y,z$ for the evolution of the spin components leading to the expression for $\frac{\text{d}\boldsymbol{S}}{\text{d}t}$.
\end{widetext}

\begin{widetext}

\section{Charge hopping processes and Lindblad operators}\label{app:4x4_to_2x2}

Sequential tunneling through a singly occupied level $\epsilon$ with a fully "$+$"-polarized left electrode occurs via two stochastic processes, each with equal probability,

\begin{align}
& \qquad
|+\rangle\langle+|
\underset{\mathrm{(discharge)}}{\overset{\mathrm{L}}{\longrightarrow}}
|\hspace{0.15em}\emptyset\hspace{0.15em}\rangle\langle\hspace{0.15em}\emptyset\hspace{0.15em}|
\underset{\mathrm{(charge)}}{\overset{\mathrm{R}}{\longrightarrow}}
|\!\uparrow\,\rangle\langle\,\uparrow\! |
\quad\leadsto\quad
\hat{L}^\mathrm{eff}_{+,\uparrow} \equiv   |\!\uparrow\,\rangle\langle\hspace{0.15em}\emptyset\hspace{0.15em}|\cdot|\hspace{0.15em}\emptyset\hspace{0.15em}\rangle\langle+| = 
|\!\uparrow\,\rangle\langle+|
\,, \label{eq:2x2_lme_1}\\
& \qquad 
|+\rangle\langle+|
\underset{\mathrm{(discharge)}}{\overset{\mathrm{L}}{\longrightarrow}}
|\hspace{0.15em}\emptyset\hspace{0.15em}\rangle\langle\hspace{0.15em}\emptyset\hspace{0.15em}|
\underset{\mathrm{(charge)}}{\overset{\mathrm{R}}{\longrightarrow}}
|\!\downarrow\,\rangle\langle\,\downarrow\! |
\quad\leadsto\quad
\hat{L}^\mathrm{eff}_{+,\downarrow} \equiv   |\!\downarrow\,\rangle\langle\hspace{0.15em}\emptyset\hspace{0.15em}|\cdot|\hspace{0.15em}\emptyset\hspace{0.15em}\rangle\langle+| = 
|\!\downarrow\,\rangle\langle+|
\,,
\label{eq:2x2_lme_2}
\end{align}

The subsequent discharge and charge processes are described by $4\times 4$ representations of $\hat{L}_n$ in Eqs.~(\ref{eq:jumps1}) to (\ref{eq:jumps4}) for different orientations of $\boldsymbol{n}_\mathrm{\alpha}$ described by $\theta,\varphi$ for both electrodes, without loss of generality. The product of Lindblad operators, however, acts only on the singly charged states and is thus reducible to $2\times 2$ representations.
This leads to effective Lindblad operators $\hat{L}^\mathrm{eff}_{f,i}$ with initial states $i\in\lbrace\uparrow,\downarrow\rbrace$ and final states $f\in\lbrace+,-\rbrace$ which act only on the spin states of the singly charged system. These yield the $2\times 2$ reduced Lindblad master equation for spin torque-driven \ac{EPR} presented in \cite{kovarik_spin_2024}. For the opposite left electrode polarization "$-$" the processes are thus given by $|\!\uparrow\,\rangle\langle-|$ and $|\!\downarrow\,\rangle\langle-|$ in the same fashion.

For sequential tunneling through the unoccupied level $\epsilon+U$, the processes are given by
\begin{align}
&\qquad 
|-\rangle\langle-|
\underset{\mathrm{(charge)}}{\overset{\mathrm{L}}{\longrightarrow}}
|\hspace{0.15em}2\hspace{0.15em}\rangle\langle\hspace{0.15em}2\hspace{0.15em}|
\underset{\mathrm{(discharge)}}{\overset{\mathrm{R}}{\longrightarrow}}
|\!\uparrow\,\rangle\langle\,\uparrow\!|
\quad\leadsto\quad
\hat{L}^\mathrm{eff}_{+,\uparrow} \equiv   |\!\uparrow\,\rangle\langle\hspace{0.15em}2\hspace{0.15em}|\cdot|\hspace{0.15em}2\hspace{0.15em}\rangle\langle-| = 
|\!\uparrow\,\rangle\langle-|
\,,\label{eq:2x2_lme_3}\\
&\qquad 
|-\rangle\langle-|
\underset{\mathrm{(charge)}}{\overset{\mathrm{L}}{\longrightarrow}}
|\hspace{0.15em}2\hspace{0.15em}\rangle\langle\hspace{0.15em}2\hspace{0.15em}|
\underset{\mathrm{(discharge)}}{\overset{\mathrm{R}}{\longrightarrow}}
|\!\downarrow\,\rangle\langle\,\downarrow\!|
\quad\leadsto\quad
\hat{L}^\mathrm{eff}_{-,\downarrow} \equiv   |\!\downarrow\,\rangle\langle\hspace{0.15em}2\hspace{0.15em}|\cdot|\hspace{0.15em}2\hspace{0.15em}\rangle\langle-| = 
|\!\downarrow\,\rangle\langle-|\,,
\label{eq:2x2_lme_4}
\end{align}

for the tip along $-$ and the $2\times 2$ dimensional Lindblad operators are $|\!\uparrow\,\rangle\langle+|$ and $|\!\uparrow\,\rangle\langle+|$ for the opposite tip "$+$" polarization.

In summary, two separate charging and discharging processes represented by the $4 \times 4$ Lindblad operators $\hat{L}_n$ in Eqs.~(\ref{eq:jumps1}) to (\ref{eq:jumps4}), yield an effective $2\times 2$ Lindblad master equation of the spin-dynamics \cite{kovarik_spin_2024}.
However, the product of Lindblad operators in Eqs.~(\ref{eq:2x2_lme_1})--(\ref{eq:2x2_lme_4}), in all four above cases, acts only on the spin-$\tfrac{1}{2}$ states and leaves the charge invariant. Thus, the product yields an effective Lindbladian representable by a $2\times2$ matrix.

\end{widetext}

\section{Second-order tunneling processes}\label{app:cotunneling}
\begin{figure}[ht]
\centering
\vspace{3em}
\begin{minipage}{1.0em}
\vspace{-8em}(a)
\end{minipage}
\begin{minipage}{11em}
\begin{fmffile}{diagram8}
  \begin{fmfgraph*}(80,40)
    \fmfstraight
    \fmfleft{i1,i2} 
    \fmfright{o1,o2}
    
    \fmf{plain_arrow,label=$\uparrow$,tension=2/5,label.side=left}{i2,v2}
    \fmf{plain,label=2,label.side=left,tension=2/5}{v2,v3}
    \fmf{plain_arrow,label=$\downarrow$,label.side=left}{v3,o2}
    
    \fmf{plain_arrow,tension=2/9,label=$\downarrow$,tension=0.25,label.side=left}{o1,v1}
    \fmf{plain,label=$\emptyset$,label.side=left}{v1,v4}
    \fmf{plain_arrow,tension=2/3,label=$\uparrow$,label.side=left}{v4,i1}
    
    \fmffreeze
    \fmf{dashes_arrow,label=L,label.side=left}{v3,v4}
    \fmf{dashes_arrow,label=R,label.side=left}{v1,v2}
    
  \end{fmfgraph*}
\end{fmffile}
\end{minipage}
\hfill
\begin{minipage}{1.0em}
\vspace{-8em}(b)
\end{minipage}
\begin{minipage}{11em}
\begin{fmffile}{diagram9}
  
  \begin{fmfgraph*}(80,40)
  
    \fmfstraight
    \fmfleft{i1,i2} 
    \fmfright{o1,o2}
    
    \fmf{plain_arrow,tension=2/5,label=$\uparrow$,label.side=left}{i2,v2}
    \fmf{plain,tension=1,label=2,label.side=left}{v2,v4}
    \fmf{plain_arrow,tension=2/3,label=$\downarrow$,label.side=left}{v4,o2}

     \fmf{plain_arrow,tension=2/5,label=$\downarrow$,label.side=left}{o1,v5}
     \fmf{plain,tension=1,label=2,label.side=left}{v5,v3}
     \fmf{plain_arrow,tension=2/3,label=$\uparrow$,label.side=left}{v3,i1}

    \fmffreeze
    \fmf{dashes_arrow,label=L,label.side=left}{v3,v2}
    \fmf{dashes_arrow,label=R,label.side=left}{v4,v5}

  \end{fmfgraph*}
\end{fmffile}
\end{minipage}

    \vspace{2em}
 \caption{Possible co-tunneling diagrams involving both virtual charging and discharging (a) and only virtual charging (b).}\label{fig:cotunneling}
\end{figure}
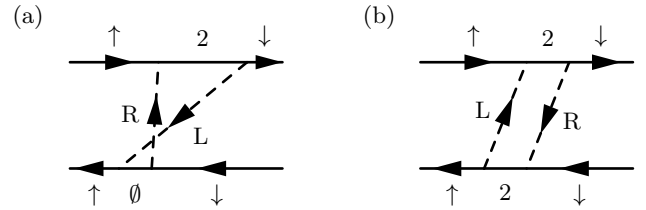

Second-order scattering processes are known as co-tunneling and are shown in Fig.~\ref{fig:cotunneling}~(a)~and~(b). All-electrical resonance driving due to co-tunneling processes~\cite{reina_galvez_cotunneling_2019} is also possible and becomes dominant where sequential tunneling is suppressed, as in the Coulomb blockaded regime~\cite{weymann_tunnel_2005}.

